\documentclass[journal]{IEEEtran}

\usepackage{amsmath,graphicx}
\usepackage{ifthen}   

\usepackage{graphicx} 
\usepackage{amssymb,amsfonts}
\usepackage{import}  % for inkscape import
\usepackage{siunitx}    % SI units
\usepackage[american]{circuitikz} % tikz circuits
\usepackage{tikz}
\usetikzlibrary{fit,matrix,chains,positioning,decorations.pathreplacing,arrows,
external}

\usepackage{fancyhdr}

\usepackage[mode=buildnew]{standalone}  % included tex files are only built if changes were made
\usepackage{pgfplots} 
\pgfplotsset{table/search path={inc},compat=1.16}
\usepackage{tikzscale}
\usepackage[section]{placeins} % for floatBarrier
\usepackage{balance}
\usepackage[normalem]{ulem}  % strike through

\newcommand{\bma}{\begin{bmatrix}}
\newcommand{\ema}{\end{bmatrix}}
\newcommand{\T}{{\mathsf{T}}} % transpose
\newcommand{\Reals}{\mathbb{R}}      % real numbers
\newcommand{\Normal}[1]{\mathcal{N}\!\left({#1}\right)} %Gaussian dist.
\newcommand{\dd}{\mathop{}\!\mathrm{d}} %d in integrals and differentiation

\newcommand{\eqdef}{\triangleq} % definition
\newcommand{\argmin}{\operatorname*{argmin}}
\newcommand{\argmax}{\operatorname*{argmax}}

\DeclareMathOperator{\E}{\textnormal{\ensuremath{\mathbb{E}}}}
\newcommand{\EE}[1]{\E\!\left[{#1}\right]}
\newcommand{\cond}{\hspace{0.02em}|\hspace{0.08em}}

\makeatletter
\DeclareFontFamily{U}{MnSymbolA}{}
\DeclareSymbolFont{MnSyA}{U}{MnSymbolA}{m}{n}
\DeclareFontShape{U}{MnSymbolA}{m}{n}{
<-6> MnSymbolA5
<6-7> MnSymbolA6
<7-8> MnSymbolA7
<8-9> MnSymbolA8
<9-10> MnSymbolA9
<10-12> MnSymbolA10
<12-> MnSymbolA12}{}
\DeclareMathSymbol{\smallrightarrow}{\mathrel}{MnSyA}{0}
\DeclareMathSymbol{\smallleftarrow}{\mathrel}{MnSyA}{2}
\DeclareMathSymbol{\smallleftrightarrow}{\mathrel}{MnSyA}{16}
\newcommand{\smallrightarrowfill@}{\arrowfill@\relbar\relbar\smallrightarrow}
\newcommand{\smallleftarrowfill@}{\arrowfill@\smallleftarrow\relbar\relbar}
\newcommand{\smallleftrightarrowfill@}
{\arrowfill@\smallleftarrow\relbar\smallrightarrow}
\renewcommand{\overrightarrow}{\mathpalette{\overarrow@\smallrightarrowfill@}}
\renewcommand{\overleftarrow}{\mathpalette{\overarrow@\smallleftarrowfill@}}
\renewcommand{\overleftrightarrow}
{\mathpalette{\overarrow@\smallleftrightarrowfill@}}
\makeatother
\providecommand{\msgf}[2]{\protect\overrightarrow{#1}_{\mspace{-3mu}#2}} % Forward Message
\newcommand{\dotl}{\begin{tabbing}...\end{tabbing}}
\newcommand{\ignore}[1]{}

\newcommand{\restrict}[2]{\left.#1\right|_{#2}}
\newcommand{\sgn}{\operatorname{sgn}}

\newcounter{examplecntr}
{\begin{trivlist}\small\item[]\refstepcounter{examplecntr}%
 {\bfseries Example~\theexamplecntr%
  \ifthenelse{\equal{#1}{}}{}{ (#1)}.
}}%
{\end{trivlist}}

\newcounter{definitioncntr}
{\begin{trivlist}\item[]\refstepcounter{definitioncntr}%
{\bfseries Definition~\thedefinitioncntr.}}%
{\hfill$\Box$\end{trivlist}}

\newcounter{theoremcntr}
\newenvironment{theorem}[1][]%
{\begin{trivlist}\item[]\refstepcounter{theoremcntr}%
{\bfseries Theorem~\thetheoremcntr%
  \ifthenelse{\equal{#1}{}}{}{ (#1)}.
}}%
{\hfill$\Box$\end{trivlist}}

\newcounter{propositioncntr}
{\begin{trivlist}\item[]\refstepcounter{propositioncntr}%
{\bfseries Proposition~\thepropositioncntr%
  \ifthenelse{\equal{#1}{}}{}{ (#1)}.
}}%
{\hfill$\Box$\end{trivlist}}
\newcounter{lemmacntr}
\newenvironment{lemma}[1][]%
{\begin{trivlist}\item[]\refstepcounter{lemmacntr}%
{\bfseries Lemma~\thelemmacntr%
  \ifthenelse{\equal{#1}{}}{}{ (#1)}.
}}%
{\hfill$\Box$\end{trivlist}}

\newcommand{\eproofnegspace}{\\[-1.5\baselineskip]\rule{0em}{0ex}}

\begin{document}

\title{A Binarizing NUV Prior and its Use for M-Level Control and Digital-to-Analog Conversion}

\author{Raphael~Keusch
        %and Hans-Andrea Loeliger,~\IEEEmembership{Fellow,~IEEE}% <-this % stops a space
        and Hans-Andrea Loeliger% <-this % stops a space
\thanks{R. Keusch and H.-A.~Loeliger are with the Department of Information 
Technology and Electrical Engineering, ETH Zurich, Zurich 8092, Switzerland
(e-mail: keusch@isi.ee.ethz.ch, loeliger@isi.ee.ethz.ch).}% <-this % stops a space
}

% make the title area
\IEEEoverridecommandlockouts
\IEEEpubid{\makebox[\columnwidth]{\tiny \shortstack{\textcopyright\ 2021 IEEE.  Personal use of this material is permitted. Permission from IEEE must be obtained for all other uses, in any\\
                                                   current or future media, including reprinting/republishing this material for advertising or promotional purposes, creating new\\
                                                   collective works, for resale or redistribution to servers or lists, or reuse of any copyrighted component of this work in other works. \hfill}} \hspace{\columnsep}\makebox[\columnwidth]{ }}
\maketitle
\IEEEpubidadjcol

\begin{abstract}
Priors with a NUV representation (normal with unknown variance)
have mostly been used for sparsity. 
In this paper, a novel NUV prior is proposed that effectively binarizes.
While such a prior may have many uses, 
in this paper, we explore its use for discrete-level control (with $M\geq 2$ levels)
including, in particular, a practical scheme for digital-to-analog conversion.
The resulting computations, for each planning period, 
amount to iterating forward-backward Gaussian message passing recursions
(similar to Kalman smoothing),
with a complexity (per iteration) that is linear in the planning horizon.
In consequence, 
the proposed method is not limited to a short planning horizon
and can therefore outperform ``optimal'' methods. 
A preference for sparse level switches can easily be incorporated. 
\end{abstract}


\IEEEpeerreviewmaketitle

%%%%%%%%%%%%%%%%%%%%%%%%%%%%%%%%%%%%%%%%%%%%%%%%%%%%%%%%%%%%%%%%%%%%%%%%%%%%%%%%%
\section{Introduction}
\label{sec:introduction}

Consider the classical control problem 
of steering an analog physical linear system along some 
desired trajectory, or to make the system produce some desired analog output signal.
In this paper, we are interested in the special case 
where the control input is discrete-level (e.g., restricted to $M$ levels),
which makes the problem much harder.
This discrete-level control problem includes, in particular, 
a certain type of digital-to-analog converter 
where the binary (or ternary) output of some digital processor directly drives 
a continuous-time analog linear filter---preferably an inexpensive one---%
which produces the desired analog waveform.

It is tempting to ask for an optimal discrete-level control signal,
i.e., a control signal that produces the best approximation 
%which is defined by its producing the best approximation 
of the desired analog trajectory 
(e.g., for a quadratic cost function). 
However, determining such an optimal control signal 
is a hard combinatorial optimization problem
with a computational complexity that is exponential in the planning horizon
%with a exponentially growing computational complexity in the planning horizon 
\cite{land_automatic_1960,wolsey_integer_1999}. 
In consequence, insisting on an optimal control signal 
effectively limits us to a short planning horizon, which is a very severe restriction.
This problem is well known in model predictive control (MPC)
\cite{aguilera_stability_2011,geyer_multistep_2014}.
Techniques such as sphere decoding do help 
\cite{dorfling_long-horizon_2019}, 
but the fundamental problem remains.

Clearly, the discrete-level input control problem is a nonconvex optimization
problem.
A general approach to nonconvex optimization 
is to resort to some convex relaxation, 
and to project the solution back to the permissible set \cite{sparrer_adapting_2014}.
Another approach uses 
ideas from sum-of-absolute-values
(SOAV) optimization~\cite{ikeda2016discrete}, 
which is an extended version of $L_1$ optimal control.
More general approaches 
include heuristic methods such as random-restart 
hill-climbing~\cite{russel_artificial_2013} and
simulated annealing~\cite{pincus_letter_1970}.

The heart of the method proposed in this paper 
is a new binarizing NUV prior, where ``NUV'' stands for ``normal with unknown variance''.
NUV priors are a central idea of sparse Bayesian learning 
\cite{tipping_sparse_2001, tipping_fast_2003,wipf_sparse_2004,wipf_new_2008},
and closely related to variational representations of 
$L_p$-norms~\cite{loeliger_factor_2018,bach_optimization_2012}.
Such priors have been used mainly for sparsity; 
in particular, no discrete-level-enforcing NUV prior seems to have been proposed
in the prior literature.
(An interesting non-NUV binarizing prior has been proposed in \cite{dai_sparse_2019}.)

A main advantage of NUV priors in general is their computational compatibility 
with linear Gaussian models, cf.~\cite{loeliger_sparsity_2016}.
In this paper, the computations (for each planning period)
amount to iterating forward-backward Gaussian message passing recursions
similar to Kalman smoothing,
with a complexity (per iteration) that is linear in the planning horizon.
In consequence, the proposed method can effectively 
handle long planning horizons, 
which can far outweigh its finding only a local minimum of the fitting cost.

The paper is organized as follows. 
In Section~\ref{sec:TwoLvlPrior},
we introduce the new NUV prior 
and demonstrate its binarizing effect in a scalar setting
with two pertinent theorems. 
In Section~\ref{sec:SystemModelAlgo},
we proceed to the binary-control problem,
and the empirical effectiveness of the proposed approach 
is demonstrated in Sections \ref{sec:Examples}--\ref{sec:OptimalController}.
In Section~\ref{sec:BeyondBinaryPriors}, we propose and demonstrate
a generalization to $M>2$ levels,
and in Section~\ref{sec:SparseSwitching}, 
we show how a preference for sparse level
switches can be easily be incorporated.

%%%%%%%%%%%%%%%%%%%%%%%%%%%%%%%%%%%%%%%%%%%%%%%%%%%%%%%%%%%%%%%%%%%%%%%%%%%%%%%%%
%\section{The Binary-Enforcing NUV Prior}
\section{The Binarizing NUV Prior}
\label{sec:TwoLvlPrior}

Let $\Normal{x; \mu, \sigma^2}$ 
denote the normal probability density function in $x$ 
with mean $\mu\in\Reals$ and variance $\sigma^2$.
Let 
\begin{equation} \label{eqn:TwoLvlPrior}
\rho(x, \theta) \eqdef \Normal{x; a, \sigma_a^2} \Normal{x; b, \sigma_b^2},
\end{equation}
where $\theta \eqdef (\sigma_a^2, \sigma_b^2)$
is a shorthand for the two variances in (\ref{eqn:TwoLvlPrior}).
The starting point of this paper is the observation
that $\rho(x, \theta)$ can be used as a (improper) joint prior for $X$ and $\theta$ 
that strongly encourages $X$ to lie in $\{ a, b \}$.

Before examining this binarizing property, 
we first note that, 
for fixed variances $\theta$,
$\rho(x, \theta)$ is a Gaussian probability density in $x$ (up to a scale
factor).
Specifically, $\rho(x, \theta)$ can be written as
\begin{equation} \label{eqn:PriorWithHyperPrior}
\rho(x, \theta) = p(x \cond \theta) \rho(\theta)
\end{equation}
with
\begin{equation} \label{eqn:ScalarPriorWithFixedTheta}
p(x \cond \theta) = \Normal{x; \mu_\theta, \sigma_\theta^2}
    %   \frac{1}{\sqrt{2\pi}\sigma_\theta} 
    %   \exp\!\left( \frac{-(x-\mu_\theta)^2}{2\sigma_\theta^2} \right)
\end{equation}
and
\begin{equation} \label{eqn:HyperPrior}
\rho(\theta) =
      \frac{1}{\sqrt{2\pi (\sigma_a^2 + \sigma_b^2)}}
      \exp\!\left( \frac{-(a-b)^2}{2(\sigma_a^2 + \sigma_b^2)} \right),
\end{equation}
where
\begin{equation} \label{eqn:PriorMean}
\mu_\theta \eqdef \frac{b\sigma_a^2 + a\sigma_b^2}{\sigma_a^2 + \sigma_b^2}
\end{equation}
and
\begin{equation}\label{eqn:PriorVariance}
\sigma_\theta^2  \eqdef  \left( 1/\sigma_a^2 + 1/\sigma_b^2 \right)^{-1}
= \frac{\sigma_a^2 \sigma_b^2}{\sigma_a^2 + \sigma_b^2}  
\end{equation}
The proof of (\ref{eqn:PriorWithHyperPrior})--(\ref{eqn:PriorVariance}) 
is given in Appendix~\ref{sec:ProofHyperPrior}.

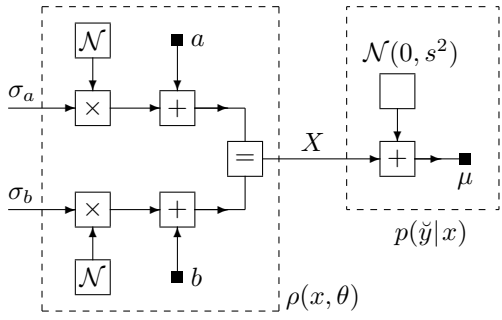
\begin{figure}[t]
\setlength{\unitlength}{0.9mm}
\newcommand{\cent}[1]{\makebox(0,0){#1}}
\newcommand{\pos}[2]{\makebox(0,0)[#1]{#2}}
\newcommand{\knownBox}{\cent{\rule{1.75\unitlength}{1.75\unitlength}}}
\newcommand{\calN}{\ensuremath{\mathcal N}}
\centering
%\small
\begin{picture}(75,45)(0,0)
%\put(0,0){\dashbox(75,45){}}

\put(0,30){\vector(1,0){10}}  \put(0,31){\pos{bl}{$\sigma_a$}}
\put(10,27.5){\framebox(5,5){$\times$}}
\put(12.5,37.5){\vector(0,-1){5}}
\put(10,37.5){\framebox(5,5){\calN}}
\put(15,30){\vector(1,0){7.5}}
\put(22.5,27.5){\framebox(5,5){$+$}}
\put(25,40){\vector(0,-1){7.5}}
\put(25,40){\knownBox}  \put(27,40){\pos{cl}{$a$}}
\put(27.5,30){\line(1,0){7.5}}
\put(27.5,30){\vector(1,0){5}}
\put(35,30){\line(0,-1){5}}
\put(0,15){\vector(1,0){10}}  \put(0,16){\pos{bl}{$\sigma_b$}}
\put(10,12.5){\framebox(5,5){$\times$}}
\put(12.5,7.5){\vector(0,1){5}}
\put(10,2.5){\framebox(5,5){\calN}}
\put(15,15){\vector(1,0){7.5}}
\put(22.5,12.5){\framebox(5,5){$+$}}
\put(25,5){\vector(0,1){7.5}}
\put(25,5){\knownBox}   \put(27,5){\pos{cl}{$b$}}
\put(27.5,15){\line(1,0){7.5}}
\put(27.5,15){\vector(1,0){5}}
\put(35,15){\line(0,1){5}}
\put(32.5,20){\framebox(5,5){$=$}}
\put(5,0){\dashbox(35,45){}}   \put(41,0){\pos{bl}{$\rho(x,\theta)$}}

\put(37.5,22.5){\vector(1,0){17.5}}  \put(45,23.6){\pos{cb}{$X$}}

\put(55,30){\framebox(5,5){}}  \put(52.5,36){\pos{bl}{$\calN(0,s^2)$}}
\put(57.5,30){\vector(0,-1){5}}
\put(55,20){\framebox(5,5){$+$}}
\put(60,22.5){\line(1,0){7.5}}
\put(60,22.5){\vector(1,0){4}}
\put(67.5,22.5){\knownBox}     \put(67.5,20.5){\pos{ct}{$\mu$}}
%
%\put(49,15){\dashbox(16,30){}}  \put(57.5,13.5){\pos{ct}{$p(\breve y \cond x)$}}
\put(50,15){\dashbox(22.5,30){}}  \put(62.5,13.5){\pos{ct}{$p(\breve y \cond x)$}}
\end{picture}
\caption{\label{fig:BinaryPriorWithGaussianLikelihood}%
Factor graph of (\ref{eqn:BinaryPriorWithGaussianLikelihood})
for fixed $\breve y$,
with parameters $\mu$ and $s^2$ depending on $\breve y$.
The boxes labeled ``\calN'' 
represent normal probability density functions $\Normal{0,1}$.
}
\end{figure}

In order to study the binarizing effect of the prior (\ref{eqn:TwoLvlPrior}),
we now assume for the rest of this section that $\rho(x, \theta)$ is used 
in some model with fixed observation(s) $\breve y$
and likelihood function $p(\breve y \cond x)$. 
Moreover, we assume $p(\breve y \cond x)$ to be Gaussian in $x$, 
with mean $\mu$ and variance $s^2$ depending on $\breve y$,
i.e.,
\begin{equation} \label{eqn:prior:GaussianLikelihood}
p(\breve y \cond x) = \gamma \Normal{x; \mu, s^2},
\end{equation}
% where ``$\propto$'' denotes equality up to a scale factor. 
where $\gamma$ is an irrelevant scale factor.
%and where the dependence of $\mu$ and $s^2$ on $\breve y$ is suppressed.
A factor graph \cite{loeliger_introduction_2004}
of the resulting statistical system model
\begin{equation} \label{eqn:BinaryPriorWithGaussianLikelihood}
p(\breve y \cond x) \rho(x, \theta)
= \gamma \Normal{x; \mu, s^2} \Normal{x; a, \sigma_a^2} \Normal{x; b, \sigma_b^2}
\end{equation}
is shown in Fig.~\ref{fig:BinaryPriorWithGaussianLikelihood}.

The detailed working of the binarizing effect of $\rho(x, \theta)$ 
depends on how the unknown variances $\theta$ are determined.
Two different ways to estimate these variances 
are considered in Sections \ref{sec:Prior:JointMAP} and~\ref{sec:Prior:VarMAP}.

\begin{figure}
\centering
\includegraphics[width=\linewidth]{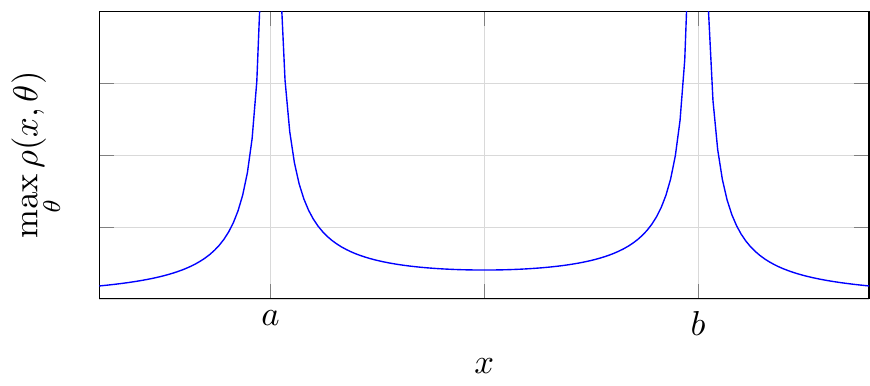}
\vspace{-6mm}

\caption{\label{fig:effectiveBinaryPrior}%
The effective (improper) prior~(\ref{eqn:ImproperPrior}) 
clearly favors $a$ and $b$.}
\end{figure}
%\vspace{\dblfloatsep}

\subsection{Joint MAP Estimation}
\label{sec:Prior:JointMAP}

\begin{figure}
\centering
\includegraphics[width=\linewidth]{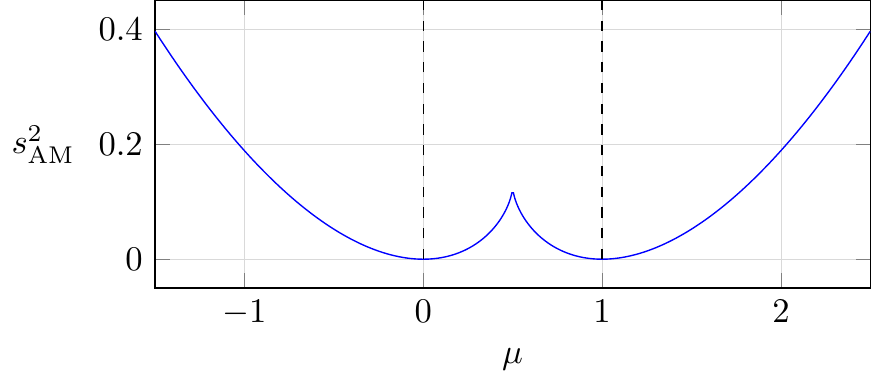}
\vspace{-6mm}

\caption{\label{fig:BoundAM}%
The value of $s^2_{\text{AM}}$ in (\ref{eqn:AM:CondNoLocal})
as a function of $\mu$
for $a=0$ and $b=1$. \\~~}
%\end{figure}
\vspace{\dblfloatsep}

%\begin{figure}
%\centering
\includegraphics[width=\linewidth]{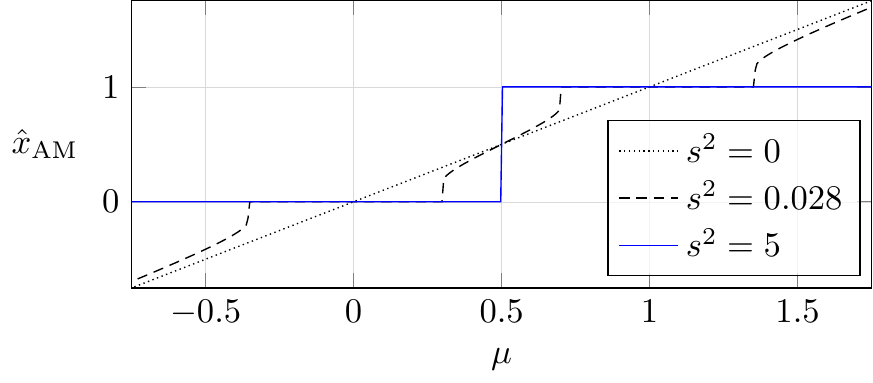}
\vspace{-6mm}

\caption{\label{fig:hatXJointMAP}%
The estimate of Section~\ref{sec:Prior:JointMAP}
for $a=0$ and $b=1$, as a function of~$\mu$.}
\end{figure}

An obvious approach to estimate $x$ and $\theta$
is by joint MAP estimation,
which results in
%Assume that $x$ and $\theta$ are determined by joint MAP estimation. 
%The resulting estimate $\hat x$ of $x$ is
\begin{IEEEeqnarray}{rCl} \label{eqn:hatXJointMAP2}
\hat x 
& = & \argmax_{x} \max_{\theta} p(\breve y \cond x) \rho(x, \theta) 
      \IEEEeqnarraynumspace\label{eqn:ScalarAM}\\
& = & \argmax_{x} 
     p(\breve y \cond x) \max_{\theta} \rho(x, \theta)
     \IEEEeqnarraynumspace\\
%& = & \argmax_{x} p(\breve y \cond x) \nonumber\\
% &&
%      \left( \max_{\sigma_a^2} \Normal{x; a, \sigma_a^2} \right) 
%      \left( \max_{\sigma_b^2} \Normal{x; b, \sigma_b^2} \right)
%      \IEEEeqnarraynumspace\\
& = & \argmax_x \frac{\Normal{x; \mu, s^2}}{|x-a|\cdot|x-b|}
      \IEEEeqnarraynumspace\label{eqn:ScalarJointMAPx}
\end{IEEEeqnarray}
where the last step follows from
%with an effective (improper) prior
\begin{IEEEeqnarray}{rCl} \label{eqn:ImproperPrior}
\max_{\theta} \rho(x, \theta)
& = & 
\max_{\sigma_a^2} \Normal{x; a, \sigma_a^2}
\max_{\sigma_b^2} \Normal{x; b, \sigma_b^2} \IEEEeqnarraynumspace\\
& \propto & \frac{1}{|x-a|\cdot |x-b|}
\label{eqn:jointMAPprior}
\end{IEEEeqnarray}
%where $\alpha$ is an irrelevant constant. 
where ``$\propto$'' denotes equality up to a scale factor. 
It is obvious that the effective prior (\ref{eqn:jointMAPprior}), 
which is plotted in Fig~\ref{fig:effectiveBinaryPrior},
has a strong preference for $x$ to lie in $\{ a, b\}$.
The following theorem guarantees that, 
%under weak conditions, 
for sufficiently large $s^2$,
the maximization in (\ref{eqn:ScalarJointMAPx})
is good-natured and returns $\hat x=a$ or $\hat x=b$.
\begin{theorem}\label{theorem:AM:scalarLocalMaxima}
The function
\begin{equation}
x \mapsto 
\frac{\Normal{x; \mu, s^2}}{|x-a|\cdot|x-b|}
\end{equation}
%(from the right-hand side of (\ref{eqn:ScalarJointMAPx}))
%(from (\ref{eqn:ScalarJointMAPx}))
has no local maximum
(other than the global maxima at $x=a$ and $x=b$) 
if and only if
\begin{equation} \label{eqn:AM:CondNoLocal}
s^2 > s^2_{\text{AM}},
\end{equation}
where $s^2_{\text{AM}}$ is the only real root
of the cubic polynomial~(\ref{eqn:AM:PolyScrit}). 
\end{theorem}
The polynomial~(\ref{eqn:AM:PolyScrit}) 
and the proof are given in Appendix~\ref{sec:TheoremScalarAM}.
Since $s^2_{\text{AM}}$ is the only real root of a cubic polynomial,
a closed-form expression for $s^2_{\text{AM}}$ exists, 
but it is cumbersome. 
However, $s^2_{\text{AM}}$ is easily computed numerically.
The value of $s^2_{\text{AM}}$ as a function of $\mu$
is plotted in Fig.~\ref{fig:BoundAM}. 
%For example, $\tilde s_{\text{AM}}^2(0.3) = 0.028$,
%% or $\mu = 0.3$ 
%in agreement with Fig.~\ref{fig:hatXJointMAP}.

In the scalar setting of this section, the estimate 
(\ref{eqn:ScalarAM}) can certainly be computed numerically for any $s^2$,
but such a brute-force approach does not generalize 
to the sequence setting of Section~\ref{sec:Algo:AM}.
%In preparation for that generalization,
With that generalization in mind,
%However, in preparation for the sequence setting of Section~\ref{sec:Algo:AM},
we now consider computing (\ref{eqn:ScalarAM}) 
%(or an approximation thereof)
by alternating maximization (AM) over $x$ and $\theta$,
which operates by alternating 
the following two steps for $i=1, 2, 3, \ldots$:
\begin{enumerate}
\item
For fixed $\theta = \theta^{(i-1)} = \left( (\sigma_a^2)^{(i-1)}, (\sigma_b^2)^{(i-1)} \right)$, 
%For fixed $\theta = \theta^{(i-1)}$, 
compute the MAP estimate
\begin{IEEEeqnarray}{rCl}
\hat x^{(i)} 
 & = & \argmax_{x} p(\breve y \cond x) \rho(x, \theta) \IEEEeqnarraynumspace\\
 & = & 
  \left( \frac{1}{\sigma_a^2} + \frac{1}{\sigma_b^2} + \frac{1}{s^2} \right)^{-1}
  \left( \frac{a}{\sigma_a^2} + \frac{b}{\sigma_b^2} + \frac{\mu}{s^2} \right).
  \IEEEeqnarraynumspace\label{eqn:scalarMAPxplicit}
\end{IEEEeqnarray}
% with $\sigma_a^2 = (\sigma_a^2)^{(i-1)}$ and $\sigma_b^2 = (\sigma_b^2)^{(i-1)}$.
%\hat x^{(i)} = 
%  \left( \frac{1}{\sigma_a^2} + \frac{1}{\sigma_b^2} + \frac{1}{s^2} \right)^{-1}
%  \left( \frac{a}{\sigma_a^2} + \frac{b}{\sigma_b^2} + \frac{\mu}{s^2} \right).
%\frac{\frac{a}{\sigma_a^2} + \frac{b}{\sigma_b^2} + \frac{\mu}{s^2}}{\frac{1}{\sigma_a^2} + \frac{1}{\sigma_b^2} + \frac{1}{s^2}}
\item
For fixed $x = \hat x^{(i)}$, compute 
\begin{equation}
\theta^{(i)} = \argmax_\theta \rho(x, \theta), 
\end{equation}
which yields
\begin{IEEEeqnarray}{rCl}
(\sigma_a^2)^{(i)} 
& = & \argmax_{\sigma_a^2} \Normal{\hat x^{(i)}; a, \sigma_a^2} \IEEEeqnarraynumspace\\
& = & \left( \hat x^{(i)} - a \right)^2
\end{IEEEeqnarray}
and likewise
\begin{equation}
(\sigma_b^2)^{(i)} = \left( \hat x^{(i)} - b \right)^2 \!.
\end{equation}
\end{enumerate}
The resulting estimate 
$\hat x_\text{AM}$ is illustrated in Fig.~\ref{fig:hatXJointMAP},
where $\sigma_a^2$ and $\sigma_b^2$ were initialized to 
$(\sigma_a^2)^{(0)}=(\sigma_b^2)^{(0)}=1$.

For general $s^2$, 
$\hat x_\text{AM}$ need not agree with (\ref{eqn:ScalarAM})
since AM may converge to a local maximum (or a saddle point).
However, if (\ref{eqn:AM:CondNoLocal}) holds,
then Theorem~\ref{theorem:AM:scalarLocalMaxima} guarantees that
AM will converge to $\hat x_\text{AM} = a$ or $\hat x_\text{AM} = b$
unless it is unluckily initialized to the (unavoidable)
local minimum between $a$ and~$b$.

\subsection{Type-II Estimation%
\protect\footnote{in the sense of \cite{tipping_sparse_2001, wipf_sparse_2004}}}
%\subsection{MAP Estimation of $\theta$}
\label{sec:Prior:VarMAP}

\begin{figure}
\centering
\includegraphics[width=\linewidth]{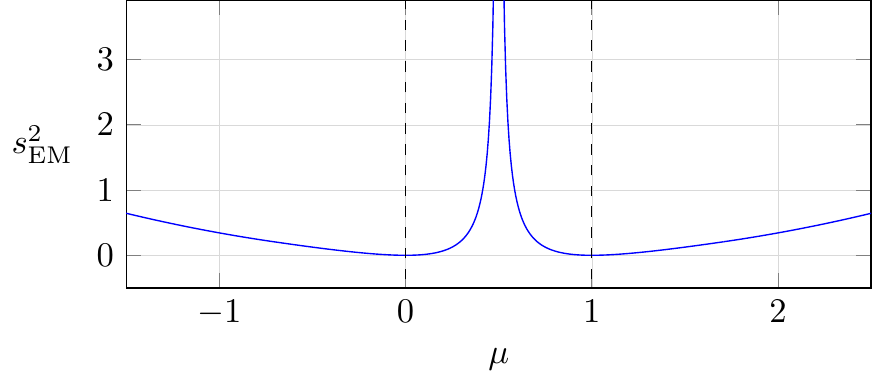}
\vspace{-6mm}

\caption{\label{fig:BoundEM}%
%Bound $\bar s^2_{\text{EM}}(\mu)$ for $a=0$ and $b=1$. 
The value of $s_{\text{EM}}^2$ in (\ref{eqn:scalarEMtheorem:Condsa}) and (\ref{eqn:scalarEMtheorem:Condsb}) 
as a function of $\mu$ for $a=0$ and $b=1$.
} 
%\end{figure}
\vspace{\dblfloatsep}

%\begin{figure}
%\centering
\includegraphics[width=\linewidth]{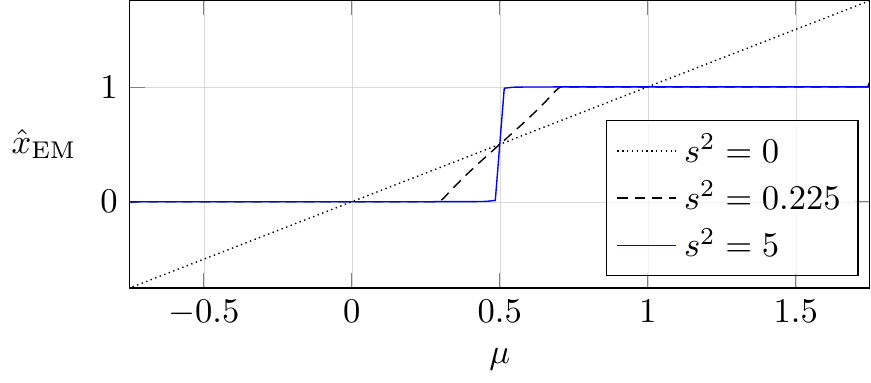}
\vspace{-6mm}

\caption{\label{fig:hatXTypeII}%
The estimate of Section~\ref{sec:Prior:VarMAP} 
for $a=0$ and $b=1$, as a function of~$\mu$.}
\end{figure}

Another approach is to first form the MAP estimate 
\begin{IEEEeqnarray}{rCl} 
\hat\theta 
%  & \eqdef & \argmax_\theta p(\theta \cond \breve y)
%       \IEEEeqnarraynumspace\\
  & = & \argmax_{\theta} \int_{-\infty}^\infty p(\breve y \cond x) \rho(x, \theta)\, \dd x,
    \IEEEeqnarraynumspace
   \label{eqn:scalarHatThetaMAP}
\end{IEEEeqnarray}
after which we estimate $x$ as
%from which we obtain the estimate
%and then we estimate $x$ as
\begin{equation} \label{eqn:hatXTypeII}
\hat x = \argmax_{x} p(\breve y \cond x) \rho(x, \hat\theta).
\end{equation}
%which is a type-II estimation of $x$ in the sense of [\ldots].
%(This is a type-II estimation of $x$ in the sense of [\ldots].)
%(We thus use type-II estimation of $x$ in the sense of [\ldots].)
Note that (\ref{eqn:hatXTypeII}) is given by (\ref{eqn:scalarMAPxplicit}).
The difference to the joint-MAP approach of Section~\ref{sec:Prior:JointMAP}
is the integration over $x$ in (\ref{eqn:scalarHatThetaMAP}).
The following theorem guarantees that, 
%under weak conditions, 
for sufficiently large $s^2$,
the maximization in (\ref{eqn:scalarHatThetaMAP})
is good-natured and (\ref{eqn:hatXTypeII}) returns $\hat x=a$ or $\hat x=b$.
\begin{theorem}\label{theorem:scalarEM}
%Let $a<b$. 
Assume $a<b$.
For $\mu < (a+b)/2$, the function
\begin{equation} \label{eqn:scalarEMtheoremIntegral}
\theta \mapsto
\int_{-\infty}^\infty \Normal{x; \mu, s^2} \rho(x,\theta)\, \dd x
\end{equation}
has a maximum at $\sigma_a^2=0$ and $\sigma_b^2=(a-b)^2$ (resulting in $\hat x = a$)
and no other extrema if and only if
\begin{IEEEeqnarray}{rCl} \label{eqn:BinCondEMa}
 s^2 > s_{\text{EM}}^2,
\end{IEEEeqnarray}
where 
%\begin{IEEEeqnarray}{rCl}
%s^2  & > & \tilde s^2_{\text{EM}} \\
% & \eqdef & 
\begin{equation}\label{eqn:scalarEMtheorem:Condsa}
s_{\text{EM}}^2 \eqdef 
    \left\{ \begin{array}{ll}
       (3-\sqrt{8})(a-\mu)(b-\mu), & \text{if~~} \mu < a - \frac{|a-b|}{\sqrt{2}} \\
       %\frac{(a-\mu)^2 (a-b)}{2\mu-(a+b)} & \text{if~~} a - \frac{|a-b|}{\sqrt{2}} \leq \mu < \frac{a+b}{2}
       \frac{(a-\mu)^2 |a-b|}{(a+b)-2\mu} & \text{if~~} a - \frac{|a-b|}{\sqrt{2}} \leq \mu < \frac{a+b}{2}
   \end{array}\right.  
   %\IEEEeqnarraynumspace\label{eqn:scalarEMtheorem:Condsa}
\end{equation}
Likewise, for $\mu > (a+b)/2$, (\ref{eqn:scalarEMtheoremIntegral})
has a maximum at $\sigma_b^2=0$ and $\sigma_a^2=(a-b)^2$ (resulting in $\hat x = b$)
and no other extrema if and only if 
\begin{IEEEeqnarray}{rCl} \label{eqn:BinCondEMb}
 s^2 > s_{\text{EM}}^2, 
\end{IEEEeqnarray}
where 
%\begin{IEEEeqnarray}{rCl}
\begin{equation}\label{eqn:scalarEMtheorem:Condsb}
%s^2  & > & %\tilde s^2_{\text{EM}} \\
 %& \eqdef & 
s_{\text{EM}}^2 \eqdef 
   \left\{ \begin{array}{ll}
       (3-\sqrt{8})(a-\mu)(b-\mu), & \text{if~~} \mu > b + \frac{|a-b|}{\sqrt{2}} \\
       \frac{(b-\mu)^2 |a-b|}{2\mu-(a+b)} & \text{if~~} \frac{a+b}{2} < \mu \leq  b + \frac{|a-b|}{\sqrt{2}} 
   \end{array}\right.
    \IEEEeqnarraynumspace
\end{equation}
%\eproofnegspace
\end{theorem}
The proof is given in Appendix~\ref{sec:TheoremScalarEM}.
The value of $s_{\text{EM}}^2$  
as a function of $\mu$ is plotted in Fig.~\ref{fig:BoundEM}.
%For example, for $\mu=0.3$, (\ref{eqn:scalarEMtheorem:Condsa}) yields 0.225,
%%$\tilde s_{\text{EM}}^2(0.3) = 0.225$,
%in agreement with Fig.~\ref{fig:hatXTypeII}.

Having in mind the generalization to the sequence setting (cf.\ Section~\ref{sec:Algo:EM}),
we now consider computing (\ref{eqn:scalarHatThetaMAP})
%(or an approximation thereof)
by expectation maximization (EM) \cite{stoica_cyclic_2004} with hidden variable $X$,
which operates by computing estimates $\theta^{(1)}, \theta^{(2)},\, \ldots\,$
according to
\begin{IEEEeqnarray}{rCl}
\theta^{(i)} & = & \argmax_{\theta} \EE{\log\!\big( p(\breve y \cond X) \rho(X, \theta) \big)\rule{0em}{2ex}}
                  \IEEEeqnarraynumspace\label{eqn:scalarThetaUpdateEM}\\
& = & \argmax_{\theta} \EE{\log \rho(X, \theta)\rule{0em}{2ex}},
      \IEEEeqnarraynumspace\label{eqn:scalarThetaUpdateEM_middle}
\end{IEEEeqnarray}
where the expectation is with respect to $p(x \cond \breve y, \theta^{(i-1)})$.
As will be detailed in Section~\ref{sec:Algo:EM}, 
the computation of (\ref{eqn:scalarThetaUpdateEM_middle})
boils down to 
\begin{equation} \label{eqn:scalarEMUpdateA}
\big( \sigma_a^2 \big)^{(i)} 
 =  V_{X}^{(i)} + \big(\hat x^{(i)} - a \big)^2
\end{equation}
and
\begin{equation} \label{eqn:scalarEMUpdateB}  
\big( \sigma_b^2 \big)^{(i)} 
 =  V_{X}^{(i)} + \big(\hat x^{(i)} -b \big)^2
\end{equation}
with $\hat x_k^{(i)}$ as in (\ref{eqn:scalarMAPxplicit})
and
\begin{equation}
V_X^{(i)} \eqdef \left( \frac{1}{\big(\sigma_a^2\big)^{(i-1)}} 
           + \frac{1}{\big(\sigma_b^2\big)^{(i-1)}} 
           + \frac{1}{s^2} \right)^{-1}\!\!\!,
\end{equation}
cf.\ the first factor in (\ref{eqn:scalarMAPxplicit}).

The estimate $\hat x_\text{EM}$ obtained by these iterations
is illustrated in Fig.~\ref{fig:hatXTypeII},
where $\sigma_a^2$ and $\sigma_b^2$ were initialized to 
$(\sigma_a^2)^{(0)}=(\sigma_b^2)^{(0)}=1$.

In general, $\hat x_\text{EM}$ need not agree with (\ref{eqn:hatXTypeII})
since EM may converge to a local maximum (or a saddle point).
However, if $\mu \neq (a+b)/2$ 
and $s^2$ satisfies (\ref{eqn:BinCondEMa}) or (\ref{eqn:BinCondEMb}),
then EM is guaranteed to converge to the desired binary solution.

%%%%%%%%%%%%%%%%%%%%%%%%%%%%%%%%%%%%%%%%%%%%%%%%%%%%%%%%%%%%%%%%%%%%%%%%%%%%%%%%%

\section{Binary Control}
\label{sec:SystemModelAlgo}

\subsection{The Problem}
\label{sec:StateSpaceProblem}

The prior (\ref{eqn:TwoLvlPrior}) may have many uses,
but in this paper, we now focus on the following application.
Consider a linear system with scalar%
\footnote{The generalization to a vector input signal is straightforward.} 
input $u_k\in\Reals$
and state $x_k\in\Reals^N$
that evolves according to 
\begin{equation} \label{eqn:StateSpaceEvolution}
x_k  = A x_{k-1} + B u_k,
%x_{k+1}  = A x_k + B u_k,   \text{\markblue{~~~check, discuss!}}
\end{equation}
where $k\in \{ 1, 2, \ldots, K\} $
%where $k\in\Integers$ 
is the time index (with finite planning horizon $K$),
and where 
both $A\in\Reals^{N\times N}$ and $B\in\Reals^{N\times 1}$ are assumed to be known.
We wish
%Our goal is 
to determine a two-level input signal 
$u_1,\, \ldots, u_K \in \{a, b\}$ 
such that some output (or feature) 
\begin{equation} \label{eqn:StateSpaceOutput}
y_k = C x_k \in \Reals^L
\end{equation}
(with known $C\in\Reals^{L\times N}$) 
follows a given target trajectory 
$\breve y_1, \ldots, \breve y_K \in\Reals^L$,
i.e., we wish
\begin{equation} \label{eqn:trajectoryCost}
\sum_{k=1}^K \|y_k - \breve y_k\|^2
\end{equation}
to be as small as possible.
The initial state $x_0$ may be known, or else it is Gaussian
as specified in Section~\ref{sec:StatModel}.

Note that this offline control
problem may represent a single episode of an online control problem with planning horizon $K$.
Note also that we are primarily interested in $K\gg 1$,
which precludes exhaustive tree search algorithms.

\subsection{Statistical Model}
\label{sec:StatModel}

\begin{figure}
\centering
% \vspace{0.5ex}
\includegraphics[width=\linewidth]{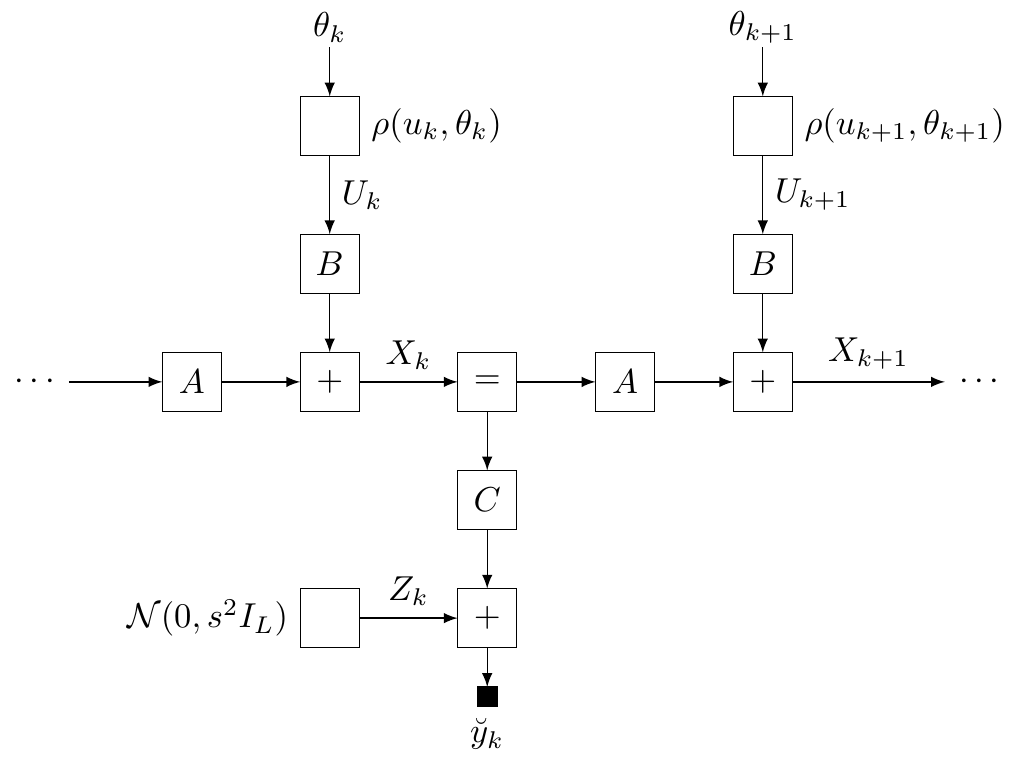} % [width=\linewidth]
% \vspace{-1ex}
\caption{\label{fig:FGSystemModel}%
Factor graph of the model~(\ref{eqn:StatModel}).}
% \end{center} 
\end{figure}

In order to solve the problem stated in Section~\ref{sec:StateSpaceProblem},
we turn it into a statistical estimation problem 
with random variables $U=(U_1,\, \ldots, U_K)$, 
$X=(X_0,\, \ldots, X_K)$, 
and with an (improper) i.i.d.\ prior%
%\footnote{Note a slight abuse of notation in (\ref{eqn:StatModelPrior}): $\rho$ is used for two different functions} 
% \footnote{In (\ref{eqn:StatModelPrior}), $\rho$ is used for two different functions (with different arguments).} 
\begin{equation} \label{eqn:StatModelPrior}
  \rho(u, \theta) \eqdef \prod_{k=1}^K \rho_k(u_k, \theta_k),
\end{equation}
where 
%$u\eqdef (u_1,\, \ldots, u_K)$,
$\theta \eqdef (\theta_1,\, \ldots, \theta_K)$
and 
\begin{equation} \label{eqn:BinSequencePrior}
\rho_k(u_k, \theta_k) \eqdef \Normal{u_k; a, \sigma_{k,a}^2} \Normal{u_k; b,
\sigma_{k,b}^2}
\end{equation}
with $\theta_k \eqdef (\sigma_{k,a}^2, \sigma_{k,b}^2)$
as in~(\ref{eqn:TwoLvlPrior}).
Accordingly, we replace (\ref{eqn:trajectoryCost}) by the likelihood function
\begin{equation} \label{eqn:LikelihoodFunction}
p(\breve y \cond u, x_0)
\eqdef \prod_{k=1}^K \frac{1}{(2\pi)^{L/2}s^L} 
                \exp\!\left( \frac{- \| y_k - \breve y_k \|^2}{2s^2} \right),
\end{equation}
where $\breve y \eqdef (\breve y_1,\, \ldots, \breve y_K)$
and where $s^2$ is a free parameter.
The initial state $X_0$ is assumed to be Gaussian 
with known mean and covariance matrix.
The complete statistical model is then given by 
%the probability law
\begin{equation} \label{eqn:StatModel}
%p(\breve y, u, x_0 \cond \theta) 
p(\breve y, u, x_0, \theta) 
\eqdef 
p(\breve y \cond u, x_0) \rho(u, \theta) p(x_0)
\end{equation}
together with (\ref{eqn:StateSpaceEvolution}) and~(\ref{eqn:StateSpaceOutput}),
cf.\ 
%and is represented as factor graph in 
Fig.~\ref{fig:FGSystemModel}.

\subsection{Iterative Kalman Input Estimation (IKIE)}
\label{sec:IKIE} \label{sec:Algorithms}

Both joint MAP estimation of $U$ and $\theta$ (as in Section~\ref{sec:Prior:JointMAP})
and type-II estimation of $U$ and $\theta$ (as in Section~\ref{sec:Prior:VarMAP}) 
can be implemented as special cases 
(with different versions of Step~2)
of the following algorithm,
which repeats the following two steps for $i=1, 2, 3,\, \ldots\, $:
\begin{enumerate}
\item
  For fixed $\theta = \theta^{(i-1)}$,
  compute the posterior means $\hat u_k^{(i)}$ of $U_k$ (for $k \in \{1,2, \ldots, K\}$) and,
  if necessary, the posterior variances $V_{U_k}^{(i)}$ of $U_k$,
  with respect to the probability distribution 
  $p(u \cond \breve y, \theta)$.
\item
  From these means and variances, determine new NUV parameters $\theta^{(i)}$:
  either by (\ref{eqn:AMUpdate_a}) and (\ref{eqn:AMUpdate_b}), 
  or by (\ref{eqn:EMUpdateA}) and (\ref{eqn:EMUpdateB}).
\end{enumerate}
Note that Step~1 operates with a standard linear Gaussian model.
In consequence, the required means and variances can be computed 
by Kalman-type recursions or, equivalently, by forward-backward Gaussian 
message passing, with a complexity that is linear in $K$.

A preferred such algorithm is MBF message passing as in \cite[Section~V]{loeliger_sparsity_2016},
which amounts to Modified Bryson–Frazier smoothing~\cite{bierman_factorization_1977}
augmented with input signal estimation. 
%this algorithm is numerically quite stable
This algorithm requires no matrix inversion%
%\footnote{This is obvious for $L=1$, but holds for every $L$.}
\footnote{This is obvious for $L=1$ and scalar input $u_k$.
For $L>1$ or vector input, a little adaptation is required.}
and is numerically quite stable. 
For the convenience of readers unfamiliar with 
the setting of \cite{loeliger_sparsity_2016},
the algorithm is concisely stated in Table~\ref{table:MBFMP}.

\begin{table}
\caption{\label{table:MBFMP}%
Step~1 of IKIE implemented by 
MBF message passing with input estimation assembled from \cite{loeliger_sparsity_2016}.}
\noindent
%\rule{\linewidth}{1pt}
\framebox[\linewidth]{\centering%
\begin{minipage}{0.9\linewidth}
\newcounter{saveequationcntr}
\setcounter{saveequationcntr}{\value{equation}}
\setcounter{equation}{0}
\renewcommand{\theequation}{M.\arabic{equation}}
%\normalsize

\noindent
\rule{0pt}{3ex}The algorithm consists of a forward recursion followed by a backward recursion.
The former is a standard Kalman filter, but the latter is not quite standard.

\noindent
\emph{\rule{0pt}{3ex}Forward recursion} for $k \in \{1,2,\,\dots,K\}$,
with $\msgf{m}{X_k}\in\Reals^{N}$ and $\msgf{V}{X_k} \in\Reals^{N\times N}$,
initialized with the mean $\msgf{m}{X_0}$ and the covariance matrix $\msgf{V}{X_0}$ 
of $X_0$ according to $p(x_0)$ in (\ref{eqn:StatModel}).
\begin{IEEEeqnarray}{rCl}
\msgf{m}{X_k} & = & 
A \Big(\msgf{m}{X_{k-1}} + \msgf{V}{X_{k-1}} C^{\T} G_{k-1} 
(\breve y_{k-1} -  C \msgf{m}{X_{k-1}}) \Big)  \nonumber \\
&& {} + B \msgf{m}{U_k} \\
\msgf{V}{X_k} &=& A F_{k-1} \msgf{V}{X_{k-1}} A^{\T} + B \msgf{V}{U_k} B^{\T} \\
\IEEEeqnarraymulticol{3}{l}{\text{with}} \nonumber \\
G_{k-1} &=& (s^2 I_L + C \msgf{V}{X_{k-1}} C^{\T})^{-1} \\
F_{k-1} &=& I_N - \msgf{V}{X_{k-1}} C^{\T} G_{k-1}  C \\
\IEEEeqnarraymulticol{3}{l}{\text{and}} \nonumber \\
  \msgf{m}{U_k} & = & 
    \mu_{\theta_k} \text{~~as in (\ref{eqn:PriorMean})}
    \IEEEeqnarraynumspace\\
  \msgf{V}{U_k} & = & 
   \sigma_{\theta_k}^2 \text{~~as in (\ref{eqn:PriorVariance}).}
\end{IEEEeqnarray}

\noindent
%\emph{\rule{0pt}{3ex}Backward recursion} 
\emph{Backward recursion} 
for $k \in \{K, K-1,\, \dots, 1\}$,
with $\tilde{\xi}_{X_k}\in\Reals^{N}$ and $\tilde{W}_{X_{k}}\in\Reals^{N\times N}$, 
initialized with 
$\tilde{\xi}_{X_{K+1}} = 0_N$ and $\tilde{W}_{X_{K+1}} = 0_{N \times N}$:
\begin{IEEEeqnarray}{rCl}
  \tilde{\xi}_{X_k} &=& F_k^{\T} A^{\T} \tilde{\xi}_{X_{k+1}} - 
  C^{\T}G_k(\breve y_k - C \msgf{m}{X_k})  \IEEEeqnarraynumspace\\
  \tilde{W}_{X_k} &=& F_k^{\T} A^{\T} \tilde{W}_{X_{k+1}} A F_k + C^{\T} G_k C.
\end{IEEEeqnarray}

\noindent
\emph{Output:} for $k\in\{ 1,2,\,\ldots,K \}$, the posterior mean is
\begin{equation}
 \hat u_k = \msgf{m}{U_k} - \msgf{V}{U_k} B^{\T} \tilde \xi_{X_k}
\end{equation}
and the posterior variance is
\begin{equation}
V_{U_k} = \msgf{V}{U_k} - \msgf{V}{U_k} B^{\T} \tilde{W}_{X_k} B \msgf{V}{U_k}.
\end{equation}

\setcounter{equation}{\value{saveequationcntr}}
\end{minipage}%
}
\end{table}

\subsection{Determining $\theta$ and $u$ by Joint MAP Estimation}
\label{sec:Algo:AM}

Joint MAP estimation of 
$U$ and $\theta$ yields
\begin{IEEEeqnarray}{rCl} \label{eqn:OptProblemJoinMAP}
\hat u 
 & = &  \argmax_{u} \max_\theta p(\breve y \cond u) \rho(u, \theta). 
         \IEEEeqnarraynumspace\label{eqn:SequenceJointMAP} \\
 & = &  \argmax_{u} p(\breve y \cond u) 
         \prod_{k=1}^K \max_{\theta_k} \rho_k(u_k, \theta_k).
        \IEEEeqnarraynumspace\label{eqn:SequenceJointMAPFactored}
\end{IEEEeqnarray}
An obvious approach to the maximization over $u$ and $\theta$ 
is to alternate between maximization over $u$ (for fixed $\theta$)
and decoupled maximizations over $\theta_1,\, \ldots, \theta_K$ (for fixed $u$).
This procedure is not guaranteed to converge to the global 
maximum, but it is very practical; 
in particular, it can be carried out by the IKIE algorithm of Section~\ref{sec:IKIE}:
Step~1 of IKIE computes the maximizing input signal 
$\hat u^{(i)} = (\hat u_1^{(i)},\, \ldots, \hat u_K^{(i)})$
while Step~2 computes the maximizing variances
\begin{equation}
\theta_k^{(i)} = \argmax_{\theta_k} \rho_k\big( \hat u_k^{(i)}, \theta_k \big)
\end{equation}
with
\begin{IEEEeqnarray}{rCl}
\big( \sigma_{k,a}^2 \big)^{(i)} 
 & = & \argmax_{\sigma_{k,a}^2} \Normal{\hat u_k^{(i)}; a, \sigma_{k,a}^2} \IEEEeqnarraynumspace\\
 & = & \big( \hat u_{k}^{(i)}\, - a \big)^2
        \label{eqn:AMUpdate_a}
\end{IEEEeqnarray}
and likewise
\begin{IEEEeqnarray}{rCl}\label{eqn:AMUpdate_b}
\big( \sigma_{k,b}^2 \big)^{(i)} &=& \big( \hat u_{k}^{(i)}\, - b \big)^2.
\end{IEEEeqnarray}

\subsection{Determining $\theta$ and $u$ by Type-II Estimation Using EM}
\label{sec:Algo:EM}

In this approach, we wish to compute the MAP estimate
%In this case, we wish to compute the estimate 
\begin{equation}\label{eqn:ThetaTypeII}
%\hat \theta = \argmax_{\theta} p(\theta \cond \breve y),
\hat \theta = \argmax_{\theta}  \int_{u} p(\breve y \cond u) \rho(u, \theta)\, \dd u.
\end{equation}
A natural approach to this maximization is expectation maximization (EM)~\cite{stoica_cyclic_2004}
with hidden variables $U$.
EM is not guaranteed to compute the global maximum,
but it results in a very practical algorithm.
Specifically, the update step for $\theta$ is 
\begin{IEEEeqnarray}{rCl}
%\theta^{(i)} & = & \argmax_{\theta} \EE{\log p(\breve y, U \cond \theta)\rule{0em}{2ex}}
%                  \IEEEeqnarraynumspace\label{eqn:ThetaUpdateEM}\\
%\theta^{(i)} & = & \argmax_{\theta} \EE{\log p(\breve y, U, \theta)\rule{0em}{2ex}}
\theta^{(i)} & = & \argmax_{\theta} \EE{\log\!\big( p(\breve y \cond U) \rho(U, \theta) \big)\rule{0em}{2ex}}
                  \IEEEeqnarraynumspace\label{eqn:ThetaUpdateEM}\\
%& = & \argmax_{\theta} \EE{\log p(\breve y \cond U) + \log \rho(U, \theta)\rule{0em}{2.3ex}}
%      \IEEEeqnarraynumspace\\
& = & \argmax_{\theta} \EE{\log \rho(U, \theta)\rule{0em}{2ex}},
      \IEEEeqnarraynumspace\label{eqn:ThetaUpdateEM_middle}
\end{IEEEeqnarray}
where the expectation is with respect to $p(u \cond \breve y, \theta^{(i-1)})$.
The maximization (\ref{eqn:ThetaUpdateEM_middle}) splits into 
\begin{IEEEeqnarray}{rCl}
\theta^{(i)}_k & = & \argmax_{\theta_k} \EE{\log \rho_k(U_k, \theta_k)\rule{0em}{2ex}},
         \IEEEeqnarraynumspace
%& = & \argmax_{\theta_k} \EE{\log \rho(U_k, \theta_k)\rule{0em}{2ex}},
%      \IEEEeqnarraynumspace
\end{IEEEeqnarray}
from which we obtain
\begin{IEEEeqnarray}{rCl}
\big( \sigma_{k,a}^2 \big)^{(i)} 
 & = & \argmax_{\sigma_{k,a}^2} \EE{\log \rule{0em}{2.3ex}\Normal{U_k; a, \sigma_{k,a}^2}}
       \IEEEeqnarraynumspace\\
 & = & \argmin_{\sigma_{k,a}^2} \left(
         \log \sigma_{k,a} 
         + \frac{1}{2\sigma_{k,a}^2}\EE{ \left( U_k - a \right)^2 }
        \right)
        \IEEEeqnarraynumspace\\
 & = &  \EE{ \left( U_k - a \right)^2 }
\end{IEEEeqnarray}
and likewise
\begin{equation}
\big( \sigma_{k,b}^2 \big)^{(i)} = \EE{ \left( U_k - b \right)^2 }.
\end{equation}
The required expectations 
can be computed by Step~1 of the IKIE algorithm of Section~\ref{sec:IKIE},
i.e.,
\begin{equation}
\EE{ U_k } = \hat u_k^{(i)}
\end{equation}
and
\begin{equation}
\EE{ U_k^2 } = V_{U_k}^{(i)} + \EE{ U_k }^2,
\end{equation}
resulting in an IKIE algorithm with Step~2 given by
\begin{equation} \label{eqn:EMUpdateA}
\big( \sigma_{k,a}^2 \big)^{(i)} 
 =  V_{U_{k}}^{(i)} + \big(\hat u_{k}^{(i)} - a \big)^2
\end{equation}
and
\begin{equation} \label{eqn:EMUpdateB}  
\big( \sigma_{k,b}^2 \big)^{(i)} 
 =  V_{U_{k}}^{(i)} + \big(\hat u_{k}^{(i)} -b \big)^2 \!.
\end{equation}

\subsection{Remarks}

\begin{enumerate}
\item
The parameter $s^2$ introduced in (\ref{eqn:LikelihoodFunction}) 
controls the approximation error (\ref{eqn:trajectoryCost}).
If $s^2$ is chosen too small, the algorithm may return a nonbinary estimate $\hat u$. 
\item
As mentioned, the algorithms of 
Sections \ref{sec:Algo:AM} and~\ref{sec:Algo:EM}
normally converge to a local (not the global) maximum 
of (\ref{eqn:SequenceJointMAP}) and (\ref{eqn:ThetaTypeII}), respectively.
However, the returned estimate $\hat u$ is often very good, 
cf.\ Section~\ref{sec:OptimalController}.
%See Section~\ref{sec:OptimalController} for a comparison with globally optimal solutions.
\item
Both versions of IKIE have the same computational complexity,
which is linear in the planning horizon $K$. 
\item
Empirically, 
type-II estimation (Section~\ref{sec:Algo:EM})
consistently outperforms joint MAP estimation (Section~\ref{sec:Algo:AM}),
which confirms what has long been know for other NUV priors \cite{giri_type_2016}.
In the numerical examples in the following sections,
only the results with type-II estimation will be reported.
\item
The generalization of (\ref{eqn:StateSpaceEvolution}) and (\ref{eqn:StateSpaceOutput})
to time-varying linear systems is obvious and has no effect on the computational cost.
Moreover, IKIE is easily adapted to mildly nonlinear systems
by adaptive linearization around the momentary trajectory estimate
in each iteration. 
\end{enumerate}

\section{Application Examples}
\label{sec:Examples}

\subsection{Digital-to-Analog Conversion}
\label{sec:ExDAC}

\begin{figure}[!t]
\begin{center}
% \vspace{0.5ex}
\includegraphics[width=\linewidth]{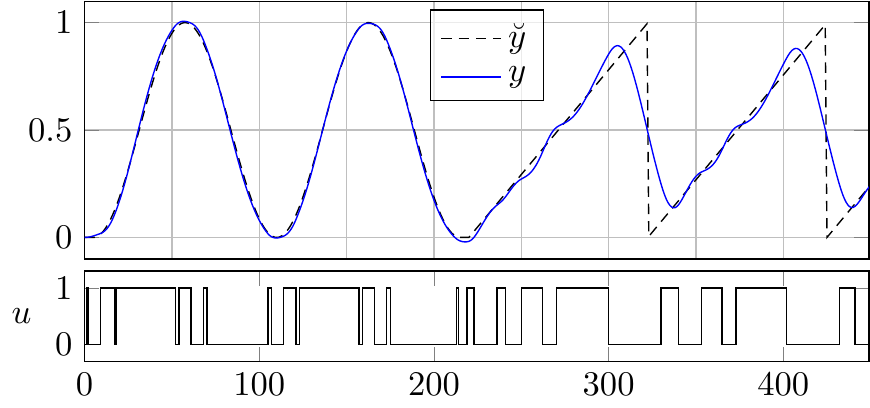} % [width=\linewidth]
% \vspace{-1ex}
\caption{\label{fig:DACWaveforms}%
Digital-to-analog conversion as in Section~\ref{sec:ExDAC} 
with target waveform $\breve y$ (dashed), digital control signal $u$ (bottom), 
and filter output signal $y$ (solid blue).}
\end{center} 
\end{figure}

One method for digital-to-analog conversion 
is to feed a continuous-time analog linear filter
directly with a binary output signal $u$ 
of a digital processor. 
This method requires an algorithm to 
compute a suitable binary signal $u$ 
such that the analog filter output approximates 
the desired analog waveform $\breve y$.
A standard approach is to compute $u$ by a delta-sigma modulator~\cite{boser_design_1988},
which requires the analog filter to approximate an ideal low-pass filter,
which may be costly.
By contrast, the method of this paper works also with much simpler 
(i.e., less expensive) analog filters.

A numerical example with such a converter is shown in
Fig.~\ref{fig:DACWaveforms}.
In this example,
the analog filter is a simple
3rd-order low-pass,   % Sallen--Key low-pass filter~\cite{sallen_practical_1955},
resulting in the discrete-time state space model 
\begin{IEEEeqnarray}{rCl}
    A &=& \bma
            0.7967&  -6.3978& -94.2123\\
0.0027 &  0.9902 & -0.1467\\
0      &  0.0030 &  0.9999
            \ema, 
 % B &=& \bma 0.0027 \\ 0 \\ 0  \ema  \quad \text{and}
 % \quad  C = \bma 0 & 0 & 35037.9 \ema           
\end{IEEEeqnarray}
$B = \bma 0.0027 & 0 & 0  \ema^{\T}\!$, and 
$C = \bma 0 & 0 & 35037.9 \ema$.
The binary input levels are $a=0$ and $b=1$.
We further have $s^2 = 0.045$ and $K=450$.

The first half of Fig.~\ref{fig:DACWaveforms} shows the normal operation of
the digital-to-analog converter, where the target trajectory 
can be well approximated.
The second half of Fig.~\ref{fig:DACWaveforms} illustrates what happens 
if the (unreasonable) target trajectory falls outside the pass band of the analog filter.

\subsection{Trajectory Planning with Sparse Checkpoints}
\label{sec:ExTrajPlanning}

\begin{figure}
\begin{center}
\includegraphics[width=\linewidth]{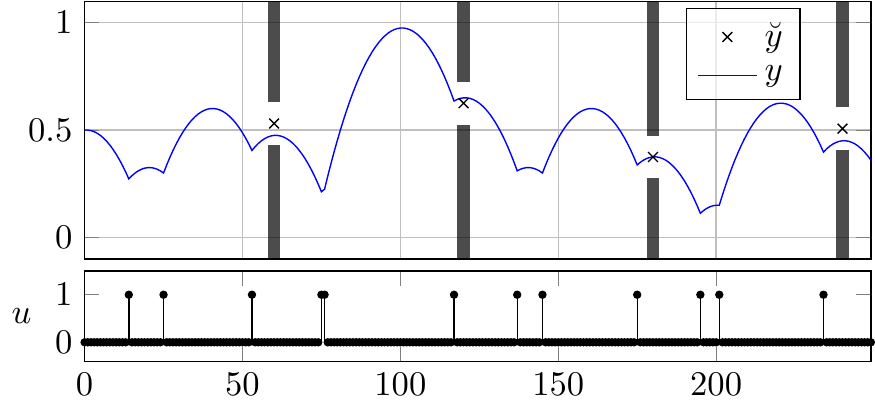}   % [width=\linewidth]
% \vspace{-0.5ex}
\caption{\label{fig:FlappyBirdTrajectory}%
Flappy bird control
with check points $\breve y$, binary control signal $u$ (bottom), 
and resulting trajectory $y$ (solid blue).}
\end{center} 
\end{figure}

The following control problem is a version of the \emph{flappy bird} computer game~\cite{wiki_flappy_bird}.
Consider an analog physical system consisting 
of a point mass $m$ moving forward (left to right in Fig.~\ref{fig:FlappyBirdTrajectory}) 
with constant horizontal velocity
and ``falling'' vertically with constant acceleration $g$. 
The $\{0,1\}$-valued control signal $u$ affects the system only if $u_k=1$,
in which case a fixed value is added to the vertical momentum.
We wish to steer the point mass such that it passes approximately 
through a sequence of check points, as illustrated in Fig.~\ref{fig:FlappyBirdTrajectory}.

For this example, we need a slight generalization%
\footnote{This generalization is effortlessly handled by IKIE.} 
of (\ref{eqn:StateSpaceEvolution})--(\ref{eqn:trajectoryCost}) as follows.
The state $x_k \in\Reals^2$ (comprising the vertical position and the vertical speed) 
evolves according to
\begin{IEEEeqnarray}{rCl}
x_k & = & \bma 1 & T \\ 0 & 1 \ema x_{k-1}
          + \bma 0 \\ 1/m \ema u_k + \bma 0 \\ -Tg \ema,
        \IEEEeqnarraynumspace
\end{IEEEeqnarray}
and we wish the vertical position 
%\begin{equation}
$y_k = \bma 1 & 0 \ema x_k$
%\end{equation}
to minimize
\begin{equation} \label{eqn:CheckpointsCost}
\sum_{k=1}^K w_k (y_k - \breve y_k)^2,
\end{equation}
where $w_k=1$ if $\breve y_k$ is a checkpoint, and $w_k=0$ otherwise, i.e., 
\begin{IEEEeqnarray}{rCl}
 w_{k} = \begin{cases} 
            1, \quad k\in\{60, 120, 180, 240\} \\ 
            0,  \quad  \text{else}. 
          \end{cases}
\end{IEEEeqnarray} 

The numerical results in Fig.~\ref{fig:FlappyBirdTrajectory}
are obtained with $m=0.5$, $T=0.1$, $g=0.25$, 
$a=0$, $b=1$, $K=250$, and $s^2 = 0.1$.

%%%%%%%%%%%%%%%%%%%%%%%%%%%%%%%%%%%%%%%%%%%%%%%%%%%%%%%%%%%%%%%%%%%%%%%%%%%%%%%%%
\section{Comparison With Other Methods}
\subsection{Exhaustive Search}
\label{sec:OptimalController}

The global minimum of (\ref{eqn:trajectoryCost})
can, in principle, be determined by an exhaustive search.
However, the complexity of such a search is exponential in the planning horizon $K$,
which limits its practicability to small $K$.
(Smart versions of tree search such as sphere decoding 
suffer from the same fundamental limitation.)

By contrast, the algorithms proposed in Section~\ref{sec:SystemModelAlgo}
will normally converge 
to a local, rather than the global, maximum of 
(\ref{eqn:SequenceJointMAP}) or (\ref{eqn:ThetaTypeII}).
However, in many applications, 
this deficiency is far outweighed by the ability to easily handle large $K$.

\begin{figure}
\begin{center}
\includegraphics[width=\linewidth]{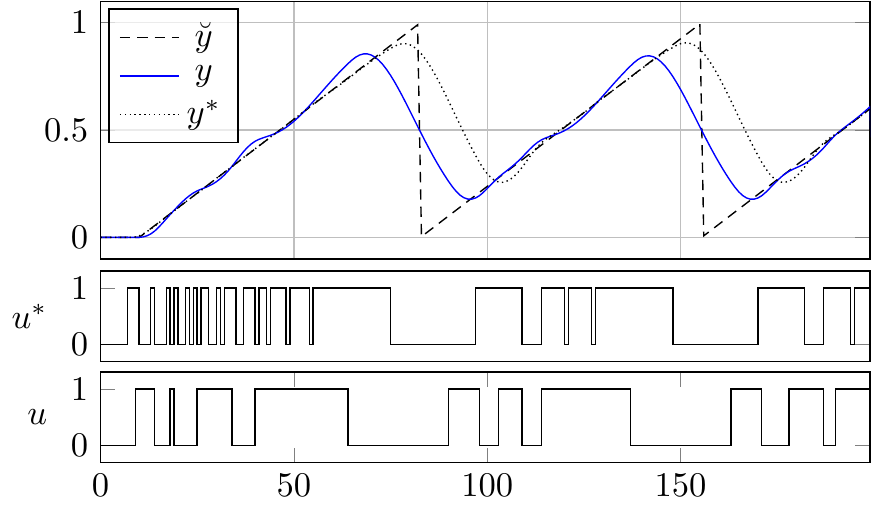}
\caption{Comparing the proposed method with an optimal (exhaustive search) controller
%with planning horizon of $K^*=8$.  
with planning horizon $K=8$.  
The former yields a significantly better approximation 
($y$ with $\text{MSE}=0.01972$) than the latter  
($y^*$ with $\text{MSE}=0.04885$).}
\label{fig:versus_optimal}
\end{center} 
%\end{figure}
%
\vspace{\floatsep}
%
%\begin{figure}
\begin{center}
\includegraphics[width=\linewidth]{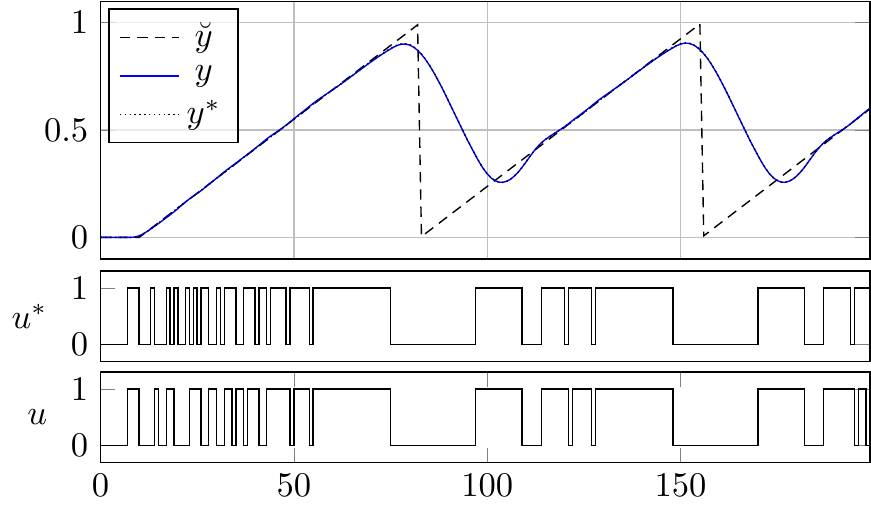}
\caption{Comparing the proposed method with planning horizon $K=8$ 
with an optimal controller with the same planning horizon. 
The former yields almost as good a solution ($y$ with $\text{MSE}=0.04899$) 
as the latter ($y^*$ with $\text{MSE}=0.04885$).}
\label{fig:versus_optimal_both_short_horizon}
\end{center} 
\end{figure}

For example, 
Fig.~(\ref{fig:versus_optimal}) compares 
an ``optimal'' (exhaustive search) controller with planning horizon $K=8$
with the proposed algorithm (IKIE with type-II estimation). 
The analog system is the same (3rd-order low-pass)
as in Section~\ref{sec:ExDAC}. 
The IKIE results are obtained with $s^2 = 0.01$ and full-length $K$.
It is obvious from Fig.~\ref{fig:versus_optimal}
that the ability to look sufficiently far ahead 
is crucial for good performance.
%outweighs the ``non-optimality'' 

But how suboptimal is the proposed algorithm really?
Fig.~\ref{fig:versus_optimal_both_short_horizon} 
shows the performance of the proposed algorithm 
in online mode with the same planning horizon $K=8$
as the exhaustive-search controller:
it turns out, in this example, that the proposed algorithm
is very nearly optimal.

\subsection{Other Ways to Gaussify Binary Variables}

Non-Gaussian variables can be approximately Gaussified
by moment matching (preferably of the posterior),
which is the basis of expectation propagation
and related methods~\cite{minka2001expectation,hu2006general}.
While such methods work well in many circumstances,
they appear to fail for the examples considered in this paper.
(Perhaps such methods have difficulties to choose among
different nearly optimal solutions.)

%%%%%%%%%%%%%%%%%%%%%%%%%%%%%%%%%%%%%%%%%%%%%%%%%%%%%%%%%%%%%%%%%%%%%%%%%%%%%%%%%

%\section{Beyond The Binary Prior}
\section{From Binary to $M$ Levels}
\label{sec:BeyondBinaryPriors}

\subsection{A False Start}

\begin{figure}
\vspace{-0.3em}
\centering
\includegraphics[width=\linewidth]{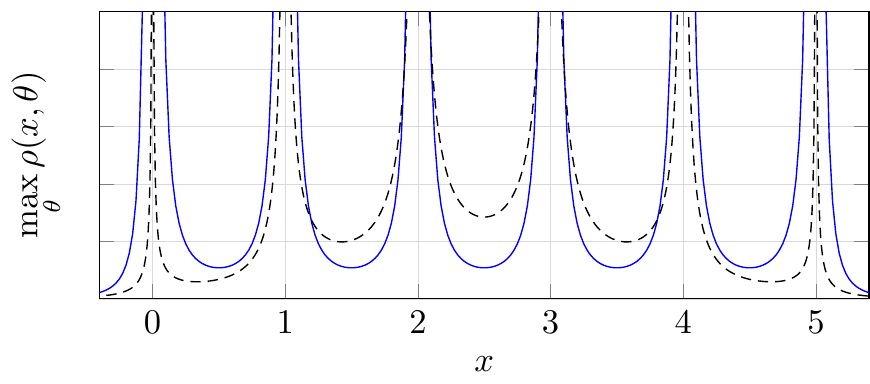}
\vspace{-6mm}

\caption{\label{fig:effectivePriorTrivialVsComposite}%
Generalization of (\ref{eqn:ImproperPrior}) to $M=6$ equidistant levels.
Solid blue: using (\ref{eqn:SumOfBinaries}) and (\ref{eqn:SumOfBinariesEqualCoeffs}).
Dashed: using (\ref{eqn:trivialPriorModel}).}
\end{figure}

An obvious attempt to generalize (\ref{eqn:TwoLvlPrior})
to more than two levels is 
\begin{equation} \label{eqn:trivialPriorModel}
\rho(x, \theta) \eqdef \Normal{x; a, \sigma_a^2} \Normal{x; b, \sigma_b^2} 
    \Normal{x; c, \sigma_c^2} \cdots
\end{equation}
with $\theta \eqdef (\sigma_a^2, \sigma_b^2,\, \ldots)$.
However, this turns out not to work very well
since it introduces a bias towards the levels in the middle range.
The effect is illustrated in Fig.~\ref{fig:effectivePriorTrivialVsComposite},
where the dashed line shows the generalization of (\ref{eqn:ImproperPrior}) 
and Fig.~\ref{fig:effectiveBinaryPrior} to 
%$\max_\theta \rho(x, \theta)$
$\rho(x, \theta)$ as in (\ref{eqn:trivialPriorModel}).

\subsection{Adding Binary Variables}

\begin{figure}
\centering
\includegraphics[width=\linewidth]{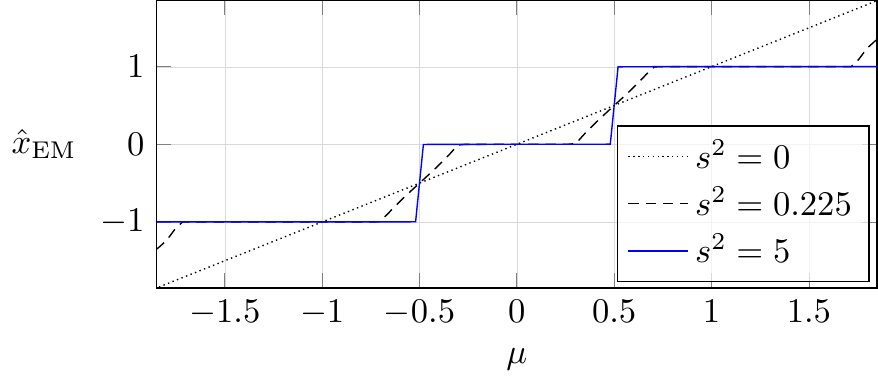}  % [width=\linewidth]
\vspace{-7mm}

\caption{\label{fig:hatXTypeIIThreeLvl}%
Generalization of Fig.~\ref{fig:hatXTypeII} to three levels $\{ -1, 0, 1 \}$
using (\ref{eqn:TernarybySumOfTwo}).}
%The estimate of Section~\ref{sec:Prior:VarMAP} as a function of $\mu$,
%where $\rho(x, \theta)$ is a three-level prior with $b=1$.}
\end{figure}

Good results are obtained with linear combinations of auxiliary binary (or binarized)
variables. For example, 
constraining $X$ to three levels $\{ -b, 0, b \}$
can be achieved by writing 
\begin{equation} \label{eqn:TernarybySumOfTwo}
X = b X_1 - b X_2
\end{equation}
where both $X_1$ and $X_2$ are constrained to $\{ 0, 1 \}$ 
by means of independent priors (\ref{eqn:TwoLvlPrior}), i.e.,
\begin{IEEEeqnarray}{rCl}
\IEEEeqnarraymulticol{3}{l}{
\rho(x_1, x_2, \theta_1, \theta_2) 
}\nonumber\\\quad
 & = & 
  \Normal{x_1; 0, \sigma_{1,a}^2} \Normal{x_1; 1, \sigma_{1,b}^2} 
  \nonumber\\
 &   &
  {}\cdot\Normal{x_2; 0, \sigma_{2,a}^2} \Normal{x_2; 1, \sigma_{2,b}^2}.
  \IEEEeqnarraynumspace
\end{IEEEeqnarray}
%with $\theta=(\ldots)$.
The corresponding generalization of Fig.~\ref{fig:hatXTypeII}
is shown in Fig.~\ref{fig:hatXTypeIIThreeLvl}.

More generally, we can write $X$ as a linear combination
\begin{equation} \label{eqn:SumOfBinaries}
X = \sum_{j=1}^J \beta_j X_j + \beta_0
\end{equation}
of independent binary (i.e., binarized to $\{ 0, 1\}$) variables $X_1,\, \ldots, X_J$.
The choice of $J$ and of the coefficients $\beta_0,\, \ldots, \beta_J$
is highly nonunique. 
Choosing $\beta_j = 2^{j-1}$ for $j>0$ does not work well empirically.
Good results are obtained with 
\begin{equation} \label{eqn:SumOfBinariesEqualCoeffs}
\beta_1 = \ldots = \beta_J,
\end{equation}
resulting in $M=J+1$ equidistant levels for $X$.
(Related representations were used in \cite{FrLg:sradda2006}.)
The corresponding generalization of (\ref{eqn:ImproperPrior})
is illustrated in Fig.~\ref{fig:effectivePriorTrivialVsComposite}.
The numerical results in the rest of this paper 
are all obtained with (\ref{eqn:SumOfBinariesEqualCoeffs}).

\subsection{Symmetry Breaking}

In (\ref{eqn:SumOfBinaries}), 
$X_1=0$ and $X_2=1$ has the same effect on $X$ as $X_1=1$ and $X_2=0$. 
The estimation algorithm must somehow 
%make a choice 
choose
among such equivalent configurations. 
However, depending on the details of the implementation,
the estimation algorithm may not, by itself, be able to break such symmetries.
This problem can be solved
by a slightly asymmetric initialization of the variances,
e.g., 
\begin{equation}
\sigma_{1,a}^2 = \sigma_{1,b}^2 \neq \sigma_{2,a}^2 = \sigma_{2,b}^2,
\end{equation}
where the inequality is almost an equality.

\subsection{Application to $M$-level Control}
\label{sec:MLevelControl}

Using (\ref{eqn:SumOfBinaries})
for the input signal of a state space model 
as in Section~\ref{sec:SystemModelAlgo}
is straightforward: 
split the input $U_k$ into independent binarized inputs $U_{k,1},\, \ldots, U_{k,J}$ 
according to
\begin{equation}
U_k = \sum_{j=1}^J \beta_j U_{k,j} + \beta_0.
\end{equation}
The corresponding modification of the state space model 
is easily handled by the IKIE algorithm.

\begin{figure}
\centering
\includegraphics[width=\linewidth]{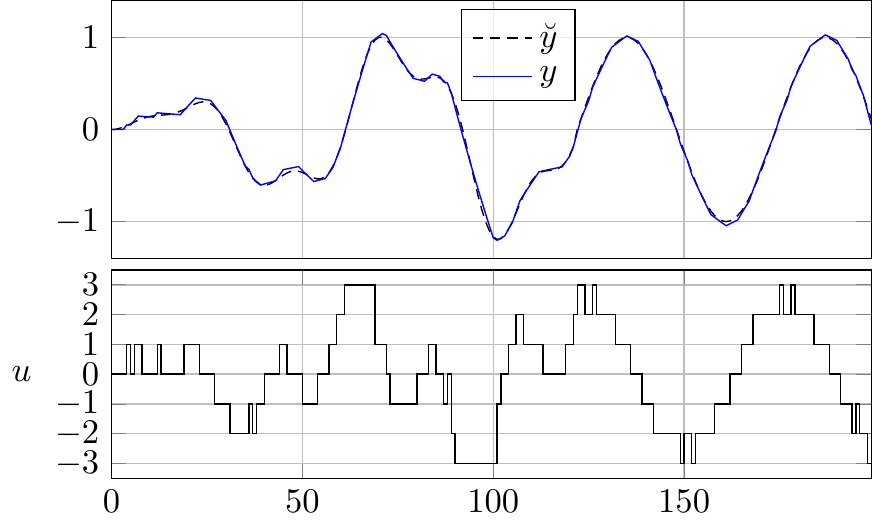}
\caption{\label{fig:largeMExample}%
The example of Section~\ref{sec:MLevelControl}:
$M$-level control with level set $\{-3,-2,1,0,1,2,3\}$,
target signal $\breve y$ (dashed), input signal $u$ (bottom), 
and resulting system output $y$ (blue).}
\end{figure}

A numerical example of $M$-level control with $M=7$
is shown in Fig.~\ref{fig:largeMExample},
where the system is a simple integrator with parameters
\begin{IEEEeqnarray}{rCl}
  A = \bma 0.98 \ema, \quad 
  B = \bma 0.05 \ema, \quad %\text{and} 
  \quad C = \bma 1 \ema,
\end{IEEEeqnarray}
$s^2=0.015$ and $K=200$.

%%%%%%%%%%%%%%%%%%%%%%%%%%%%%%%%%%%%%%%%%%%%%%%%%%%%%%%%%%%%%%%%%%%%%%%%%%%%%%%%%
\section{Sparse Level Switching}
\label{sec:SparseSwitching}

In some applications, switching the input signal 
between the allowed discrete levels is costly 
(e.g., because of thermal losses in electronic power switches) 
and should be done as infrequently as possible. 
Fortunately, the approach of Section~\ref{sec:SystemModelAlgo} 
can easily be adapted to address this additional requirement.

\subsection{The Method}

The key idea is to extend the state space model such 
that the level differences $u_k - u_{k-1}$ 
appear as an additional output signal to which a power penalty can be applied.
For binary control, 
this is achieved by extending 
(\ref{eqn:StateSpaceEvolution}) and (\ref{eqn:StateSpaceOutput})
to
\begin{equation}
\tilde x_k = \tilde A_{k-1} \tilde x_{k-1} + \tilde B u_k
\end{equation}
and
\begin{equation}
\tilde y_k = \tilde C \tilde x_k
\end{equation}
with 
\begin{IEEEeqnarray}{rClrCl}
\label{eqn:augmentation1}
  \tilde A &=& \bma 
          \begin{array}{c|c}
            \begin{matrix} 
              0 & 0  \\
              1 & 0  \\
            \end{matrix}
          & 0_{2 \times N}\\
          \hline \\[-1em]
           0_{N \times 2}  & A
          \end{array}
        % 0 & 0 & \\
        % 1 & 0 & \\
        % & & \bar A_k
      \ema, 
  \quad  
  & \tilde B &=& 
        \left [
        \begin{array}{c}
          \begin{matrix}
            1 \\
            0 \\
          \end{matrix} \\
          \hline \\[-1em]
          B 
          \end{array}
        \right ], \quad \text{and} \IEEEeqnarraynumspace  \\
  \label{eqn:augmentation2}
  \tilde C &=& \bma 
            \begin{array}{c|c}
                \begin{matrix} 1 & -1 \end{matrix}  & 0_{1 \times N} \\
                  \hline \\[-1em]
                   0_{1 \times N} &  C
            \end{array}
        \ema,
\end{IEEEeqnarray}
resulting in the extended output signal
\begin{equation} \label{eqn:InputDifferenceAsOutput}
\tilde y_k = \tilde C \tilde x_k = (u_k - u_{k-1}, y_k).
\end{equation}
(For $M$-level control as in Section~\ref{sec:MLevelControl},
the matrix $\tilde B$ is easily modified 
so that (\ref{eqn:InputDifferenceAsOutput}) holds also in this case.)
We then extend the target signal $\breve y$ to
\begin{equation} \label{eqn:OutputWithInputDifference}
\mathring y \eqdef (0, \breve y)
\end{equation}
and the likelihood function (\ref{eqn:LikelihoodFunction}) to
\begin{IEEEeqnarray}{rCl} 
%\IEEEeqnarraymulticol{3}{l}{
p(\mathring{y} \cond u, \tilde x_0)
%}\nonumber\\\quad
& \eqdef &        
\prod_{k=1}^K \frac{1}{(2\pi)^{L/2} s^L} 
                \exp\!\left( \frac{- \| y_k - \breve y_k \|^2}{2s^2} \right)
       \IEEEeqnarraynumspace\nonumber\\
 & & {}\cdot \frac{1}{\sqrt{2\pi} \tilde s} 
          \exp\!\left( \frac{-\tilde y_{k,1}^2}{ 2\tilde s^2} \right),
     \label{eqn:LikelihoodExtendedSparsity}
\end{IEEEeqnarray}
where $\tilde y_{k,1} = u_k - u_{k-1}$ denotes 
the first component of (\ref{eqn:OutputWithInputDifference}),
$\tilde x_0 = (0, 0, x_0^\T)^\T$, 
and where $\tilde s^2$ is a free parameter which controls the sparsity 
level of the input signal.

Note that the IKIE algorithm of Section~\ref{sec:IKIE} 
is easily adapted to handle this extended model.

\begin{figure}
\begin{center}
\includegraphics[]{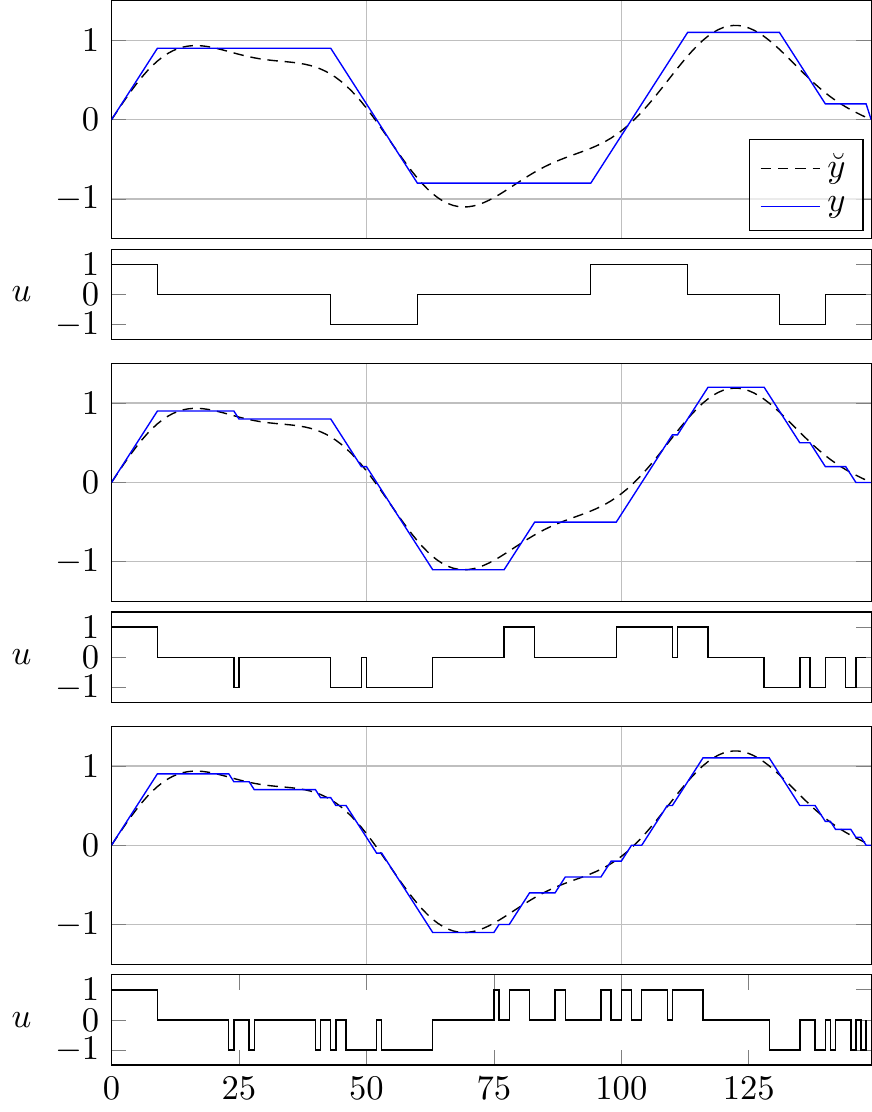}
\caption{\label{fig:SparseLvlSwitchesMotorExample}%
Motor control as in Section~\ref{sec:ExElMotorControl} with target
$\breve y$ (dashed), three-level control signal $u$, and resulting 
motor coil current $y$ (solid blue). 
Top: $s^2=4$ and $\tilde s^2=10$; 
middle: $s^2=1$ and $\tilde s^2=100$; 
bottom: $s^2=0.01$ and $\tilde s^2=10000$.
}
\end{center} 
\end{figure}

\subsection{An Example}
\label{sec:ExElMotorControl}

A numerical example with this method is shown 
in Fig.~\ref{fig:SparseLvlSwitchesMotorExample}.
The system model is an integrator with
\begin{IEEEeqnarray}{rCl}
\label{eqn:motor_lssm}
%   A =  \bma 1 \ema , \quad  B =  \bma 1/\alpha \ema  \quad \text{and} \quad 
   A =  \bma 1 \ema , \quad  B =  \bma 0.1 \ema  \quad \text{and} \quad 
   C= \bma 1 \ema,
\end{IEEEeqnarray}
which may be viewed as a simple model%
---an inductor controlled by a voltage input---%
of an electric motor,
where the state variable that we wish to control is the current through the coil.

% ignore

%%%%%%%%%%%%%%%%%%%%%%%%%%%%%%%%%%%%%%%%%%%%%%%%%%%%%%%%%%%%%%%%%%%%%%%%%%%%%%%%%
\section{Conclusion}
\label{sec:Conclusion}

We have introduced a new binarizing NUV prior
and demonstrated its use for binary and $M$-level control
and digital-to-analog conversion. 
A preference for sparse level switches can easily be incorporated.
The actual computations are iterations of Kalman-type forward-backward recursions, 
with a complexity (per iteration) that is linear in the planning horizon.
This linear complexity compares favorably with existing ``optimal'' methods.
The suitability of the binarizing prior for other applications remains to be investigated.

%%%%%%%%%%%%%%%%%%%%%%%%%%%%%%%%%%%%%%%%%%%%%%%%%%%%%%%%%%%%%%%%%%%%%%%%%%%%%%%%%
\appendices

\section{Proof of (\ref{eqn:PriorWithHyperPrior})--(\ref{eqn:PriorVariance})}
\label{sec:ProofHyperPrior}

The claim is that
\begin{IEEEeqnarray}{rCl}
\rho(x, \theta) 
 & = & \Normal{x; a, \sigma_a^2} \Normal{x; b, \sigma_b^2} \IEEEeqnarraynumspace\\
 % & = & \frac{1}{2\pi \sigma_a\sigma_b}
 %      \exp\!\left( \frac{-(x-\mu_\theta)^2}{2\sigma_\theta^2} \right)
 %      \exp\!\left( \frac{-(a-b)^2}{2(\sigma_a^2 + \sigma_b^2)} \right)
 %      %\IEEEeqnarraynumspace\label{eqn:PriorExpressionMainStep}\\
 %      \nonumber\\
 % &&   \label{eqn:PriorExpressionMainStep}\\
 & = & \frac{1}{\sqrt{2\pi}\sigma_\theta} 
       \exp\!\left( \frac{-(x-\mu_\theta)^2}{2\sigma_\theta^2} \right)
       \IEEEeqnarraynumspace\nonumber\\
 &  &  \frac{1}{\sqrt{2\pi (\sigma_a^2 + \sigma_b^2)}}
       \exp\!\left( \frac{-(a-b)^2}{2(\sigma_a^2 + \sigma_b^2)} \right).
       \IEEEeqnarraynumspace
\end{IEEEeqnarray}
The correctness of the exponents follows from
\begin{IEEEeqnarray}{rCl}
\IEEEeqnarraymulticol{3}{l}{
\frac{(x-a)^2}{2\sigma_a^2} + \frac{(x-b)^2}{2\sigma_b^2} - \frac{(x-\mu_\theta)^2}{2\sigma_\theta^2} 
}\nonumber\\\quad
& = & 
  x^2 \left( \frac{1}{2\sigma_a^2} + \frac{1}{2\sigma_b^2} - \frac{1}{2\sigma_\theta^2} \right)
  - x \left( \frac{a}{\sigma_a^2} + \frac{b}{\sigma_b^2} - \frac{\mu_\theta}{\sigma_\theta^2}\right)
      \IEEEeqnarraynumspace\nonumber\\
  &&  {} + \frac{a^2}{2\sigma_a^2} + \frac{b^2}{2\sigma_b^2} - \frac{\mu_\theta^2}{2\sigma_\theta^2}
      \IEEEeqnarraynumspace\\
& = & \frac{a^2}{2\sigma_a^2} + \frac{b^2}{2\sigma_b^2} 
      - \frac{(b\sigma_a^2 + a\sigma_b^2)^2}{2(\sigma_a^2 + \sigma_b^2) \sigma_a^2 \sigma_b^2}
      \IEEEeqnarraynumspace\\
& = & \frac{a^2 (\sigma_a^2 + \sigma_b^2) \sigma_b^2 + b^2 (\sigma_a^2 + \sigma_b^2) \sigma_a^2
       - (b\sigma_a^2 + a\sigma_b^2)^2}%
      {2(\sigma_a^2 + \sigma_b^2) \sigma_a^2 \sigma_b^2}
      \IEEEeqnarraynumspace\\
& = & \frac{(a^2 -2ab +b^2) \sigma_a^2 \sigma_b^2}%
       {2(\sigma_a^2 + \sigma_b^2) \sigma_a^2 \sigma_b^2}
      \IEEEeqnarraynumspace\\
& = & \frac{(a-b)^2}{2(\sigma_a^2 + \sigma_b^2)}
\end{IEEEeqnarray}
The correctness of the prefactors follows from
\begin{equation}
\sigma_\theta \sqrt{\sigma_a^2 + \sigma_b^2} = \sigma_a \sigma_b.
\end{equation}

\section{Proof of Theorem~\ref{theorem:AM:scalarLocalMaxima}}
\label{sec:TheoremScalarAM}

\begin{figure}
\centering
\includegraphics[]{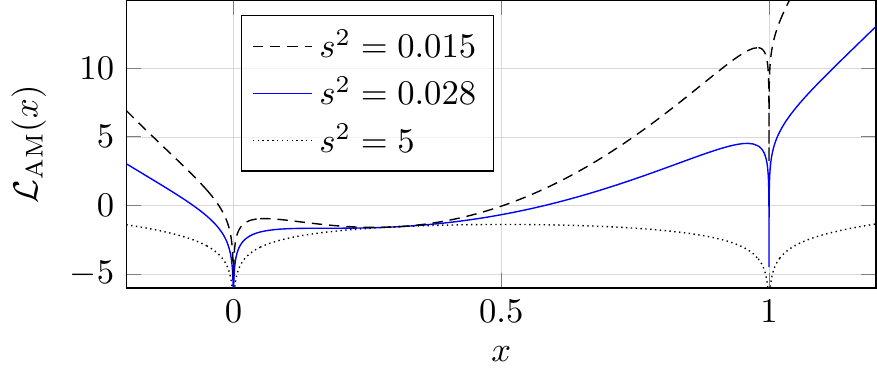}
\vspace{-6mm}

\caption{The cost function (\ref{eqn:CostFunctionAM}) for $a=0$, $b=1$, 
$\mu = 0.3$, and different values of $s^2.$ For small $s^2$ (dashed), 
an additional local minimum appears.}
%which is undesired.}
\label{fig:CostFunctionAM}
\end{figure}

We examine 
\begin{IEEEeqnarray}{rCl}
\mathcal{L}_{\text{AM}}(x) 
 & \eqdef & - \log \frac{\Normal{x; \mu, s^2}}{|x-a|\cdot|x-b|} 
       \IEEEeqnarraynumspace\\
 & = & \frac{(\mu - x)^2}{2 s^2}  + \log|x - a| +  \log|x - b| \nonumber \\
    && + \frac{\log(2 \pi s^2 )}{2} ,
        \IEEEeqnarraynumspace\label{eqn:CostFunctionAM}
\end{IEEEeqnarray}
which is illustrated in Fig.~\ref{fig:CostFunctionAM}.
Clearly, for any $s^2 > 0$, 
the singularities at $x=a$ and $x=b$ are global minima of (\ref{eqn:CostFunctionAM}).
Between these two singularities,
by continuity, there is at least one local maximum,
and there may be additional local minima and maxima.

The derivative of (\ref{eqn:CostFunctionAM}) is
%Extremal points of $\mathcal L_{AM}(x)$ are found by setting the derivative 
\begin{IEEEeqnarray}{rCl} \label{eqn:CostAM:Derivative}
  \frac{\partial }{\partial x} \mathcal L_\text{AM}(x) 
  % = - \frac{1}{s^2}(\mu - x) + \frac{1}{x-a} + \frac{1}{x-b}
   = \frac{1}{s^2}(x - \mu) + \frac{1}{x-a} + \frac{1}{x-b}
   \IEEEeqnarraynumspace
\end{IEEEeqnarray}
In order to examine the existence of such additional local minima and maxima,
we set (\ref{eqn:CostAM:Derivative}) to zero,
which yields
\begin{IEEEeqnarray}{rCl}
 c_3 x^3 + c_2 x^2 + c_1 x + c_0  = 0  
   \IEEEeqnarraynumspace\label{eqn:PolyCoeffAM}
\end{IEEEeqnarray}
with coefficients
%\begin{IEEEeqnarray}{rClcrCl}
\begin{IEEEeqnarray}{rCl}
%  c_0 &=& s^2 (-a -b) -a b \mu, \IEEEyesnumber\IEEEyessubnumber* \\ % & \quad &
  c_0 &=& s^2 (-a -b) -a b \mu \\
  c_1 &=& 2 s^2 + a b + (a + b) \mu \IEEEeqnarraynumspace \\
  c_2 &=& -a -b -\mu \\
  c_3 &=& 1. 
\end{IEEEeqnarray}
% https://en.wikipedia.org/wiki/Cubic_equation
The left-hand side of (\ref{eqn:PolyCoeffAM}) is a cubic polynomial
with discriminant
%the left-hand side of (\ref{eqn:PolyCoeffAM}) is
%defined as
\begin{equation}\label{eqn:ProofAM:Discriminant}
  \Delta  =  -27 c_3^2 c_0^2 + 18 c_3 c_2 c_1 c_0 - 4 c_3 c_1^3 - 4 c_2^3 c_0 + c_2^2 c_1^2.  
  \IEEEeqnarraynumspace
\end{equation}
If $\Delta > 0$, then (\ref{eqn:PolyCoeffAM}) has three real solutions, 
and if $\Delta < 0$, then (\ref{eqn:PolyCoeffAM}) has only one real solution.
(These two cases are illustrated in Fig.~\ref{fig:CostFunctionAM}
by the dashed line and the dotted line, respectively.)
Recall that there is at least one local maximum between $a$ and $b$.
Thus (\ref{eqn:CostFunctionAM}) has no local minimum if and only if
$\Delta < 0$.

We proceed to examine the condition $\Delta < 0$ as a function of $s^2$.
The following calculations are very cumbersome 
and are preferably carried out (or verified) with the aid of computer algebra.
We begin by writing (\ref{eqn:ProofAM:Discriminant}) as
\begin{equation} \label{eqn:ProofAM:DeltaOfs2}
\Delta = \phi(s^2),
\end{equation}
where $\phi$ is the polynomial
\begin{equation} \label{eqn:AM:PolyScrit}
\phi(\zeta) \eqdef \tilde c_3 \zeta^3 + \tilde c_2 \zeta^2 + \tilde c_1 \zeta + \tilde c_0
\end{equation}
with coefficients
\begin{IEEEeqnarray}{rCl}
\tilde c_0 
% & = & {} - 27 (ab\mu)^2 + 18(a+b+\mu)ab\mu (ab + a\mu + b\mu)^2 \nonumber\\
% && {} - 4(ab + a\mu + b\mu)^3  - 4(a+b+\mu)^3 ab\mu \nonumber\\
% && {} + (a+b+\mu)^2 (ab + a\mu + b\mu)^2   %\text{~~(??)}
%   \IEEEeqnarraynumspace\\
 & = & (a-b)^2 (a-\mu)^2 (b-\mu)^2 \IEEEeqnarraynumspace\\
\tilde c_1 
% & = &  -54 (a+b)ab\mu \nonumber\IEEEeqnarraynumspace\\
% && {} + 18(a+b+\mu)\Big( (a+b)(ab+a\mu+b\mu) + 2ab\mu \Big) \nonumber\IEEEeqnarraynumspace\\
% && {} - 24(ab+a\mu+b\mu) -4(a+b+\mu)^3 (a+b) \nonumber\\
% && {} + 4(a+b+\mu)^2 (ab+a\mu+b\mu)  %\text{~~(??)}
%   \IEEEeqnarraynumspace\\
 & = & -2(a-b)^2 \Big( 2(a^2+b^2) +ab -5\mu(a+b-\mu) \Big)
   \IEEEeqnarraynumspace\\
\tilde c_2 
% & = & -27(a+b)^2 + 36(a+b+\mu)(a+b) \nonumber\\
% && {} - 48(ab+a\mu+b\mu) + 4(a+b+\mu)^2  %\text{~~(ok)}
%   \IEEEeqnarraynumspace\\
 & = & 13(a-b)^2 + 4(ab-a\mu-b\mu+\mu^2)
       \IEEEeqnarraynumspace\\
\tilde c_3 & = & -32.    %\text{~~(ok)}
   \IEEEeqnarraynumspace
\end{IEEEeqnarray}

Since 
%$\tilde c_0>0$ and $\tilde c_3 < 0$, (\ref{eqn:ProofAM:DeltaOfs2}) is positive for $s^2=0$ and 
$\tilde c_3 < 0$, (\ref{eqn:ProofAM:DeltaOfs2}) is 
negative for sufficiently large $s^2$.
%Moreover, (\ref{eqn:AM:PolyScrit}) \ldots
%
It remains to show that (\ref{eqn:AM:PolyScrit})
has exactly one real root.
But the discriminant of (\ref{eqn:AM:PolyScrit}) is
\begin{IEEEeqnarray}{rCl}
\tilde\Delta 
 & = & -27 \tilde c_3^2 \tilde c_0^2 + 18 \tilde c_3 \tilde c_2 \tilde c_1 \tilde c_0 
   - 4 \tilde c_3 \tilde c_1^3 - 4 \tilde c_2^3 \tilde c_0 + \tilde c_2^2 \tilde c_1^2
   \IEEEeqnarraynumspace\\
 & = & -16(a-b)^2 (a+b-2\mu)^4 \gamma^3
\end{IEEEeqnarray}
with
\begin{equation}
%\gamma \eqdef 7(a-b)^2 + ab - (a+b)\mu + \mu^2.
\gamma \eqdef 6(a-b)^2 + a^2 + b^2 + \mu^2 - (ab + a\mu + b\mu).
\end{equation}
For $a\neq b$, 
$\gamma > 0$ 
since $ab < (a^2 + b^2)/2$, $a\mu \leq (a^2 + \mu^2)/2$, and $b\mu \leq (b^2 + \mu^2)/2$.
Thus $\tilde \Delta < 0$, which implies that (\ref{eqn:AM:PolyScrit})
has exactly one real root.

\section{Proof of Theorem~\ref{theorem:scalarEM}}
\label{sec:TheoremScalarEM}

\subsection{Rewriting the Function}

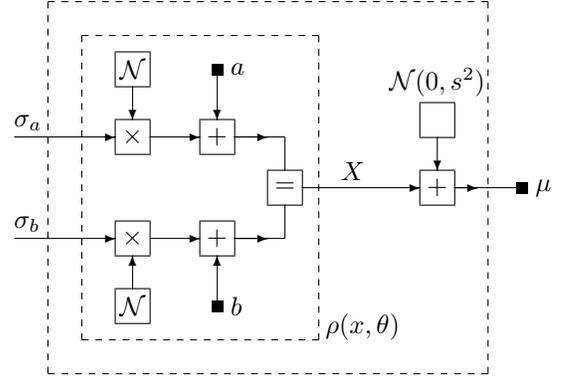
\begin{figure}
\setlength{\unitlength}{0.9mm}
\newcommand{\cent}[1]{\makebox(0,0){#1}}
\newcommand{\pos}[2]{\makebox(0,0)[#1]{#2}}
\newcommand{\knownBox}{\cent{\rule{1.75\unitlength}{1.75\unitlength}}}
\newcommand{\calN}{\ensuremath{\mathcal N}}
\centering
%\small
\begin{picture}(80,55)(-5,-5)
%\put(-5,-5){\framebox(80,55){}}

\put(-5,30){\vector(1,0){15}}  \put(-5,31){\pos{bl}{$\sigma_a$}}
\put(10,27.5){\framebox(5,5){$\times$}}
\put(12.5,37.5){\vector(0,-1){5}}
\put(10,37.5){\framebox(5,5){\calN}}
\put(15,30){\vector(1,0){7.5}}
\put(22.5,27.5){\framebox(5,5){$+$}}
\put(25,40){\vector(0,-1){7.5}}
\put(25,40){\knownBox}  \put(27,40){\pos{cl}{$a$}}
\put(27.5,30){\line(1,0){7.5}}
\put(27.5,30){\vector(1,0){5}}
\put(35,30){\line(0,-1){5}}
\put(-5,15){\vector(1,0){15}}  \put(-5,16){\pos{bl}{$\sigma_b$}}
\put(10,12.5){\framebox(5,5){$\times$}}
\put(12.5,7.5){\vector(0,1){5}}
\put(10,2.5){\framebox(5,5){\calN}}
\put(15,15){\vector(1,0){7.5}}
\put(22.5,12.5){\framebox(5,5){$+$}}
\put(25,5){\vector(0,1){7.5}}
\put(25,5){\knownBox}   \put(27,5){\pos{cl}{$b$}}
\put(27.5,15){\line(1,0){7.5}}
\put(27.5,15){\vector(1,0){5}}
\put(35,15){\line(0,1){5}}
\put(32.5,20){\framebox(5,5){$=$}}
\put(5,0){\dashbox(35,45){}}   \put(41,0){\pos{bl}{$\rho(x,\theta)$}}

\put(37.5,22.5){\vector(1,0){17.5}}  \put(45,23.6){\pos{cb}{$X$}}

\put(55,30){\framebox(5,5){}}  \put(57.5,36){\pos{cb}{$\calN(0,s^2)$}}
\put(57.5,30){\vector(0,-1){5}}
\put(55,20){\framebox(5,5){$+$}}
\put(60,22.5){\line(1,0){10}}
\put(60,22.5){\vector(1,0){3.5}}
\put(70,22.5){\knownBox}     \put(72,22.5){\pos{cl}{$\mu$}}
%
%\put(49,15){\dashbox(16,30){}}  \put(57.5,13.5){\pos{ct}{$p(\breve y \cond x)$}}

\put(0,-5){\dashbox(65,55){}}  %\put(66,-5){\pos{bl}{$\propto p(\mu \cond \theta)$}}
\end{picture}
\caption{\label{fig:ScalarEMFunctionIntegral}%
The step from (\ref{eqn:ScalarEMFunctionIntegrable}) 
to (\ref{eqn:ScalarEMFunctionIntegrated}).
}
\end{figure}

Using (\ref{eqn:PriorWithHyperPrior}) and (\ref{eqn:ScalarPriorWithFixedTheta}),
the function (\ref{eqn:scalarEMtheoremIntegral})
can be written as 
\begin{IEEEeqnarray}{rCl}
%\mathcal{L}_{\text{EM}}(\theta) 
% & \eqdef & - \log \int_{-\infty}^\infty \Normal{x; \mu, s^2} \rho(x,\theta)\, dx
\IEEEeqnarraymulticol{3}{l}{
\int_{-\infty}^\infty \Normal{x; \mu, s^2} \rho(x,\theta)\, \dd x
}\nonumber\\\quad
 & = & \rho(\theta) \int_{-\infty}^\infty \Normal{x; \mu, s^2} p(x \cond \theta)\, \dd x
       \IEEEeqnarraynumspace\\
 & = & \rho(\theta) \int_{-\infty}^\infty \Normal{\mu; x, s^2} \Normal{x; \mu_\theta, \sigma_\theta}\, \dd x
       \IEEEeqnarraynumspace\label{eqn:ScalarEMFunctionIntegrable}\\
 & = & \rho(\theta) \Normal{\mu - \mu_\theta; \sigma_\theta^2 + s^2},
       \label{eqn:ScalarEMFunctionIntegrated}
\end{IEEEeqnarray}
cf.\ Fig.~\ref{fig:ScalarEMFunctionIntegral}.
Inserting (\ref{eqn:HyperPrior}), 
taking logarithms, changing the sign, and dropping irrelevant constants
yields
\begin{IEEEeqnarray}{rCl}
\mathcal{L}_{\text{EM}}(\theta) 
 & \eqdef &  \log(\sigma_a^2 + \sigma_b^2) + \frac{(a-b)^2}{\sigma_a^2 + \sigma_b^2} \nonumber\\
 && {} + \log(\sigma_\theta^2 + s^2) + \frac{(\mu-\mu_\theta)^2}{\sigma_\theta^2 + s^2} 
        \IEEEeqnarraynumspace\label{eqn:DefCostScalarEM}
\end{IEEEeqnarray}
Note that the minima and maxima of (\ref{eqn:DefCostScalarEM})
are exactly the maxima and minima, respectively, 
of (\ref{eqn:scalarEMtheoremIntegral}).

For 
$\sigma_a^2>0$ and $\sigma_b^2>0$,
(\ref{eqn:DefCostScalarEM}) is continuous and differentiable 
both in $\sigma_a^2$ and in $\sigma_b^2$.
Moreover, 
$\lim_{\sigma_a^2 + \sigma_b^2 \rightarrow 0} \mathcal{L}_{\text{EM}}(\theta) = \infty$.
In consequence, 
%as a function of $\sigma_a^2$ and $\sigma_b^2$,
(\ref{eqn:DefCostScalarEM}) has at least one minimum 
for finite $\sigma_a^2\geq 0$ and $\sigma_b^2  \geq 0$.

In the following, we will occasionally use the assumption $a<b$
from the theorem.

\subsection{Derivatives}

\begin{IEEEeqnarray}{rCl}
\frac{\partial \mathcal{L}_{\text{EM}}(\theta)}{\partial \sigma_a^2}
& = & \frac{1}{\sigma_a^2 + \sigma_b^2}
    - \frac{(a-b)^2}{(\sigma_a^2 + \sigma_b^2)^2}
    + \frac{1}{\sigma_\theta^2 + s^2} \cdot \frac{\partial \sigma_\theta^2}{\partial \sigma_a^2}
   \IEEEeqnarraynumspace \nonumber\\
& & {} + \frac{2(\mu_\theta-\mu)}{\sigma_\theta^2 + s^2} \cdot \frac{\partial \mu_\theta}{\partial \sigma_a^2}
    - \frac{(\mu_\theta-\mu)^2}{(\sigma_\theta^2 + s^2)^2} \cdot \frac{\partial \sigma_\theta^2}{\partial \sigma_a^2}
    \label{eqn:negLLDerSigmaa}
\end{IEEEeqnarray}
with
\begin{IEEEeqnarray}{rCl}
\frac{\partial \mu_\theta}{\partial \sigma_a^2}
& = & \frac{b}{\sigma_a^2 + \sigma_b^2} - \frac{b\sigma_a^2 + a\sigma_b^2}{(\sigma_a^2 + \sigma_b^2)^2}
      \IEEEeqnarraynumspace\\
& = & \frac{(b-a)\sigma_b^2}{(\sigma_a^2 + \sigma_b^2)^2}
\end{IEEEeqnarray}
and
\begin{IEEEeqnarray}{rCl}
\frac{\partial \sigma_\theta^2}{\partial \sigma_a^2}
& = & \left(\frac{\sigma_b^2}{\sigma_a^2 + \sigma_b^2}\right)^2.
     \IEEEeqnarraynumspace\label{eqn:PriorVarDerSigmaa}
%& = & (1 + \sigma_a^2 / \sigma_b^2)^{-2}.
%     \IEEEeqnarraynumspace
\end{IEEEeqnarray}
By symmetry, 
$\partial \mathcal{L}_{\text{EM}}(\theta) / \partial \sigma_b^2$ 
is obtained by exchanging $a$ and $b$, and $\sigma_a^2$ and $\sigma_b^2$,
in (\ref{eqn:negLLDerSigmaa})--(\ref{eqn:PriorVarDerSigmaa}).

\subsection{Binarizing Minima}

For $\sigma_b^2=0$, (\ref{eqn:negLLDerSigmaa}) becomes 
\begin{equation}
\frac{\partial \mathcal{L}_{\text{EM}}(\theta)}{\partial \sigma_a^2} = 
\frac{\sigma_a^2 - (a-b)^2}{\sigma_a^4}
\end{equation}
It follows that the point 
\begin{equation} \label{eqn:BinarizingPoint}
\sigma_b^2=0  \text{~~and~~}  \sigma_a^2 = (a-b)^2 
\end{equation}
is a minimum of (\ref{eqn:DefCostScalarEM}) if and only if 
$\partial \mathcal{L}_{\text{EM}}(\theta) / \partial \sigma_b^2 > 0$.
By symmetry, the point
\begin{equation} \label{eqn:BinarizingPoint2}
%\sigma_b^2 = (a-b)^2 \text{~~and~~} \sigma_a^2=0
\sigma_a^2=0 \text{~~and~~} \sigma_b^2 = (a-b)^2 
\end{equation}
is a minimum of (\ref{eqn:DefCostScalarEM}) if and only if 
$\partial \mathcal{L}_{\text{EM}}(\theta) / \partial \sigma_a^2 > 0$.
We now examine this condition.

At the point (\ref{eqn:BinarizingPoint2}), 
we have $\mu_\theta=a$, $\sigma_\theta^2=0$, $\partial \sigma_\theta^2 / \partial \sigma_a^2 = 1$,
and (\ref{eqn:negLLDerSigmaa}) becomes
\begin{IEEEeqnarray}{rCl}
\IEEEeqnarraymulticol{3}{l}{
\frac{\partial \mathcal{L}_{\text{EM}}(\theta)}{\partial \sigma_a^2}
 = \frac{1}{(a-b)^2} - \frac{1}{(a-b)^2} + \frac{1}{s^2}
  }\nonumber\\\quad
 &   & {} + \frac{2(a-\mu)}{s^2} \cdot
       \frac{(b-a)}{(a-b)^2} - \frac{(a-\mu)^2}{s^4}
       \IEEEeqnarraynumspace\\ 
 & = & \frac{1}{s^2} \left(
        1 + \frac{2(a-\mu)}{b-a} - \frac{(a-\mu)^2}{s^2}
        \right)
       \IEEEeqnarraynumspace\\
 & = & \frac{2}{s^2 (b-a)} \left(
         \frac{a+b}{2} - \mu - \frac{(b-a)(a-\mu)^2}{2 s^2}
        \right)
\IEEEeqnarraynumspace
\end{IEEEeqnarray}
Recalling the assumption $b>a$,
it follows that (\ref{eqn:BinarizingPoint2})
is a minimum of (\ref{eqn:DefCostScalarEM})
if and only if 
\begin{equation} \label{eqn:muCond4a}
\mu < \frac{a+b}{2} - \frac{(b-a)(a-\mu)^2}{2 s^2}
\end{equation}
or, equivalently,
if and only if $\mu < (a+b)/2$ and
\begin{equation} \label{eqn:s2Cond4a}
s^2 > \frac{(b-a)(a-\mu)^2}{a+b-2\mu}
\end{equation}
By symmetry, (\ref{eqn:BinarizingPoint})
is a minimum of $\mathcal{L}_{\text{EM}}(\theta)$ if and only if
\begin{equation} \label{eqn:muCond4b}
\mu > \frac{a+b}{2} + \frac{(b-a)(b-\mu)^2}{2 s^2}
\end{equation}
or, equivalently,
if and only if $\mu > (a+b)/2$ and
\begin{equation} \label{eqn:s2Cond4b}
s^2 > \frac{(b-a)(b-\mu)^2}{2\mu - (a+b)}
\end{equation}

\subsection{Other Extrema --- Part~I}

The rest of the proof is about excluding any other extrema,
i.e., extrema with $\sigma_a^2 > 0$ and $\sigma_b^2 > 0$.
Such extrema are characterized by the conditions
\begin{equation} \label{eqn:ProofEMDerSigmaaZero}
\frac{\partial \mathcal{L}_{\text{EM}}(\theta)}{\partial \sigma_a^2} = 0.
\end{equation}
and
\begin{equation} \label{eqn:ProofEMDerSigmabZero}
\frac{\partial \mathcal{L}_{\text{EM}}(\theta)}{\partial \sigma_b^2} = 0.
\end{equation}

The following calculations are very cumbersome 
and are preferably carried out (or verified) with the aid of computer algebra.
%In preparation for analyzing these conditions,
We begin by eliminating $\mu_\theta$ and $\sigma_\theta^2$
in (\ref{eqn:negLLDerSigmaa}) 
using (\ref{eqn:PriorMean}) and (\ref{eqn:PriorVariance}), 
i.e.,
\begin{equation}
\mu_\theta - \mu = \frac{b\sigma_a^2 + a\sigma_b^2 - \mu(\sigma_a^2 + \sigma_b^2)}{\sigma_a^2 + \sigma_b^2}
\end{equation}
and
\begin{equation}
\sigma_\theta^2 + s^2 
= \frac{\sigma_a^2 \sigma_b^2 + s^2 \sigma_a^2 + s^2 \sigma_b^2}{\sigma_a^2 + \sigma_b^2}
\end{equation}
We thus obtain
\begin{IEEEeqnarray}{rCl}
\IEEEeqnarraymulticol{3}{l}{
\frac{\partial \mathcal{L}_{\text{EM}}(\theta)}{\partial \sigma_a^2}
= \Big( \sigma_a^2 \sigma_b^2 + s^2 \sigma_a^2 + s^2 \sigma_b^2 \Big)^{-2}
}\nonumber\\\quad
 & & {}\cdot\Big( \sigma_a^2 (s^2 + \sigma_b^2)^2 - \sigma_b^4 (a-\mu)^2 - s^4(a-b)^2
        \IEEEeqnarraynumspace\nonumber\\
  & & {~~~} - 2s^2 \sigma_b^2 (a^2 - a\mu + b\mu - ab) + s^2 \sigma_b^2(s^2+\sigma_b^2) \Big).
        \IEEEeqnarraynumspace\label{eqn:ProofEM:Der1}
\end{IEEEeqnarray}
%and the analogous expression for 
%$\frac{\partial \mathcal{L}_{\text{EM}}(\theta)}{\partial \sigma_a^2}$
%is obtained by exchanging $\sigma_a^2$ and $\sigma_b^2$, and $a$ and $b$.
Note that (\ref{eqn:ProofEM:Der1}) 
is a fraction of two polynomials 
%in $\sigma_a^2$ and $\sigma_b^2$,
with a strictly positive denominator and a numerator that is linear in $\sigma_a^2$.
We can thus solve (\ref{eqn:ProofEMDerSigmaaZero}) for $\sigma_a^2$,
resulting in
\begin{IEEEeqnarray}{rCl}
\sigma_a^2 & = & (s^2 + \sigma_b^2)^{-2} 
        \Big( \sigma_b^4 (a-\mu)^2 + s^4(a-b)^2
        \IEEEeqnarraynumspace\nonumber\\
  && {} + 2s^2 \sigma_b^2 (a^2 - a\mu + b\mu - ab) - s^2 \sigma_b^2(s^2+\sigma_b^2)
        \Big).
       \IEEEeqnarraynumspace
\end{IEEEeqnarray}
Likewise, setting (\ref{eqn:ProofEMDerSigmabZero}) to zero yields
\begin{IEEEeqnarray}{rCl}
\sigma_b^2 & = & (s^2 + \sigma_a^2)^{-2} 
        \Big( \sigma_a^4 (b-\mu)^2 + s^4(a-b)^2
        \IEEEeqnarraynumspace\nonumber\\
  && {} + 2s^2 \sigma_a^2 (b^2 - b\mu + a\mu - ab) - s^2 \sigma_a^2(s^2+\sigma_a^2)
        \Big).
       \IEEEeqnarraynumspace\label{eqn:ProofEM:sigma2FromDeriv}
\end{IEEEeqnarray}
Inserting (\ref{eqn:ProofEM:sigma2FromDeriv}) into (\ref{eqn:ProofEM:Der1})
yields
\begin{equation} \label{eqn:ProofEMDerSigmaaZeroSep}
%\frac{\partial \mathcal{L}_{\text{EM}}(\theta)}{\partial \sigma_a^2}
\restrict{\frac{\partial \mathcal{L}_{\text{EM}}(\theta)}{\partial \sigma_a^2}}{(\ref{eqn:ProofEMDerSigmabZero})}
= \frac{A_a\sigma_a^4 + B_a\sigma_a^2 + C_a}{(s^2 + \sigma_a^2)^2 \big( s^2(b-a) + \sigma_a^2 (b-\mu) \big)}
\end{equation}
with coefficients
\begin{IEEEeqnarray}{rCl}
A_a & = & b - \mu \label{eqn:ProofEMDerSigmaaZeroSepCoeffA}\\
B_a & = & s^2(a+2b-3\mu) - (a-\mu)^2 (b-\mu)
        \IEEEeqnarraynumspace\\
C_a & = & s^4(a+b-2\mu) + s^2 (a-\mu)^2 (a-b).
        \IEEEeqnarraynumspace\label{eqn:ProofEMDerSigmaaZeroSepCoeffC}
\end{IEEEeqnarray}
By symmetry, we likewise obtain
\begin{equation} \label{eqn:ProofEMDerSigmabZeroSep}
%\frac{\partial \mathcal{L}_{\text{EM}}(\theta)}{\partial \sigma_b^2}
\restrict{\frac{\partial \mathcal{L}_{\text{EM}}(\theta)}{\partial \sigma_b^2}}{(\ref{eqn:ProofEMDerSigmaaZero})}
= \frac{A_b\sigma_b^4 + B_b\sigma_b^2 + C_b}{(s^2 + \sigma_b^2)^2 \big( s^2(a-b) + \sigma_b^2 (a-\mu) \big)}
\end{equation}
with coefficients
\begin{IEEEeqnarray}{rCl}
A_b & = & a - \mu \\
B_b & = & s^2(b+2a-3\mu) - (b-\mu)^2 (a-\mu)
        \IEEEeqnarraynumspace\\
C_b & = & s^4(a+b-2\mu) + s^2 (b-\mu)^2 (b-a).
        \IEEEeqnarraynumspace
\end{IEEEeqnarray}

The point of these manipulations is that 
(\ref{eqn:ProofEMDerSigmaaZeroSep}) does not depend on $\sigma_b^2$
and (\ref{eqn:ProofEMDerSigmabZeroSep}) does not depend on $\sigma_a^2$,
and we have
\begin{lemma} \label{lemma:ProofEM:Separated}
Eq.\ (\ref{eqn:ProofEMDerSigmaaZero}) and (\ref{eqn:ProofEMDerSigmabZero})
hold simultaneously if 
% (???) 
% \markblue{\\to check!!!\\}
and only if  
(\ref{eqn:ProofEMDerSigmaaZeroSep}) and (\ref{eqn:ProofEMDerSigmabZeroSep})
are both zero.
\end{lemma}

\subsection{Other Extrema --- Part~II}

The numerator of (\ref{eqn:ProofEMDerSigmaaZeroSep}) is $\phi_a(\sigma_a^2)$
with
\begin{equation} \label{eqn:ProofEMDeraPoly}
\phi_a(\zeta) \eqdef A_a \zeta^2 + B_a \zeta + C_a.
\end{equation}
%has a positive real root.
We thus need to examine the conditions for (\ref{eqn:ProofEMDeraPoly})
to have a positive real root.
Clearly, (\ref{eqn:ProofEMDeraPoly}) has real roots
if and only if
\begin{equation} \label{eqn:ProofEMDerSigmaaZeroSepDiscr}
B_a^2 - 4A_aC_a \geq 0.
\end{equation}
Plugging in%
\footnote{again preferably done or verified by computer algebra} 
(\ref{eqn:ProofEMDerSigmaaZeroSepCoeffA})--(\ref{eqn:ProofEMDerSigmaaZeroSepCoeffC})
turns (\ref{eqn:ProofEMDerSigmaaZeroSepDiscr}) into
\begin{equation} \label{eqn:ProofEMDerSigmaaZeroSepDiscr2}
%(a-\mu)^2 \Big( s^4 -6(a - \mu)(b - \mu) s^2  +  (a-\mu)^2 (b - \mu)^2 \Big)
(a-\mu)^2 \psi(s^2) \geq 0
\end{equation}
with
\begin{equation}
\psi(\xi) \eqdef \xi^2 -6(a - \mu)(b - \mu) \xi  +  (a-\mu)^2 (b - \mu)^2.
\end{equation}
It is easily verified that $\psi(\xi)=0$ if and only if
\begin{equation}
\xi = |a-\mu|\cdot |b-\mu| \left( 3\sgn(a-\mu)\sgn(b-\mu) \pm \sqrt{8} \right) 
\end{equation}
where $\sgn$ denotes the sign function
\begin{equation}
\sgn(z) \eqdef \left\{ \begin{array}{ll}
         +1, & \text{if $z\geq 0$} \\
         -1, & \text{if $z<0$.}
      \end{array}\right.
\end{equation}
If $a< \mu  < b$ 
%(and recalling the assumption $a<b$),
the zeros of $\psi$ are negative,
which implies $\psi(s^2)\geq 0$ for all $s^2$.
For $\mu\leq a$ or $\mu\geq b$, the zeros of $\psi$ are positive
and $\psi(s^2) < 0$ if and only if 
\begin{equation} \label{eqn:ProofEM:NoRealRootsIntervalS}
\big( 3 - \sqrt{8} \big) (a-\mu)(b-\mu) < s^2 < \big( 3 + \sqrt{8} \big) (a-\mu)(b-\mu).
\end{equation}
By symmetry, the same conditions apply also for the numerator of 
(\ref{eqn:ProofEMDerSigmabZeroSep}).
In summary, we have
\begin{lemma} \label{lemma:ProofEM:ComplexZeros}
The polynomial (\ref{eqn:ProofEMDeraPoly})
has no real zeros% 
---i.e., (\ref{eqn:ProofEMDerSigmaaZeroSepDiscr}) does not hold---%
if and only if
(\ref{eqn:ProofEM:NoRealRootsIntervalS}) holds.
The same condition applies also to the numerator of (\ref{eqn:ProofEMDerSigmabZeroSep}).
\end{lemma}

We now assume that (\ref{eqn:ProofEMDerSigmaaZeroSepDiscr}) holds,
and we examine the conditions for (\ref{eqn:ProofEMDeraPoly}) 
to have at least one positive root. 
If $A_a>0$, then the condition is $-B_a + \sqrt{B_a^2 - 4A_a C_a} \geq 0$;
if $A_a<0$, then the condition is $-B_a - \sqrt{B_a^2 - 4A_a C_a} \leq 0$.
These two conditions boil down to
\begin{equation}
A_a B_a \leq 0 \text{~~or~~} A_a C_a \leq 0.
\end{equation}
We thus obtain
\begin{lemma}\label{lemma:ProofEM:NoPositiveRoots1}
Assume that (\ref{eqn:ProofEMDerSigmaaZeroSepDiscr}) holds.
Then (\ref{eqn:ProofEMDerSigmaaZeroSep}) has no positive real zero
if and only if
\begin{equation} \label{eqn:ProofEMNoPos1}
A_a B_a > 0 \text{~~and~~} A_a C_a > 0,
\end{equation}
and (\ref{eqn:ProofEMDerSigmabZeroSep}) has no positive real zero
if and only if
\begin{equation} \label{eqn:ProofEMNoPos2}
A_b B_b > 0 \text{~~and~~} A_b C_b > 0.
\end{equation}
\eproofnegspace
\end{lemma}

\begin{table*}
\caption{\label{table:EMProof:curves}%
Ordering of critical functions in the proof of Theorem~\ref{theorem:scalarEM},
cf.\ Fig.~\ref{fig:EMProof:curves}.}
\vspace{-3mm}
\[
\begin{array}{c|c}
\text{range} & \text{order}\\\hline
\rule{0em}{4ex}%
\mu < a - \frac{b-a}{\sqrt{2}} &
      \frac{(a-\mu)^2(b-a)}{a+b-2\mu}
    < (3-\sqrt{8})(a-\mu)(b-\mu) 
    < \frac{(a-\mu)^2(b-\mu)}{a+2b-3\mu}
    < (3+\sqrt{8})(a-\mu)(b-\mu) 
    \\
\rule{0em}{4ex}%
a - \frac{b-a}{\sqrt{2}}  < \mu < a &
      \frac{(a-\mu)^2(b-\mu)}{a+2b-3\mu}
    < \frac{(a-\mu)^2(b-a)}{a+b-2\mu}
    < (3-\sqrt{8})(a-\mu)(b-\mu) 
    < (3+\sqrt{8})(a-\mu)(b-\mu)
    \\
\rule{0em}{4ex}%
a < \mu < \frac{a+b}{2} &
      (3+\sqrt{8})(a-\mu)(b-\mu)
    < (3-\sqrt{8})(a-\mu)(b-\mu)
    < \frac{(a-\mu)^2(b-\mu)}{a+2b-3\mu}
    < \frac{(a-\mu)^2(b-a)}{a+b-2\mu}
\end{array}
\]
\end{table*}

We further examine the terms in (\ref{eqn:ProofEMNoPos1}).
%and (\ref{eqn:ProofEMNoPos2}).
The condition $A_a B_a > 0$ expands to
\begin{equation} \label{eqn:ProofEM:CondAaBa}
\left\{ \begin{array}{ll}
s^2 > \frac{(a-\mu)^2 (b-\mu)}{a+2b-3\mu} 
     & \text{if $\mu<\frac{a+2b}{3}$ or $\mu > b$} 
     \IEEEeqnarraynumspace\\
\text{false,} & \text{otherwise.}
            \IEEEeqnarraynumspace
\end{array} \right.
\end{equation}
The condition $A_a C_a>0$ expands to
\begin{equation} \label{eqn:ProofEM:CondAaCa}
\left\{ \begin{array}{ll}
s^2 > \frac{(a-\mu)^2 (b-a)}{a+b-2\mu}, 
     & \text{if $\mu < \frac{a+b}{2}$} 
     \IEEEeqnarraynumspace\\
\text{false,} & \text{if $\frac{a+b}{2} \leq \mu \leq b$}
            \IEEEeqnarraynumspace\\
\text{true,} & \text{if $\mu > b$.}
         \IEEEeqnarraynumspace
\end{array} \right.
\end{equation}
We also note that the expressions
$\frac{(a-\mu)^2 (b-\mu)}{a+2b-3\mu}$
and $\frac{(a-\mu)^2 (b-a)}{a+b-2\mu}$
as functions of $\mu$
intersect at the three points $\mu=a$ and
\begin{equation}\label{eqn:ProofEM:muInsection}
\mu = a \pm \frac{b-a}{\sqrt{2}}
\end{equation}
cf.\ Table~\ref{table:EMProof:curves}.
From (\ref{eqn:ProofEM:CondAaBa})--(\ref{eqn:ProofEM:muInsection}), we can conclude that 
(\ref{eqn:ProofEMNoPos1}) is equivalent to
\begin{equation} \label{eqn:ProofEM:CondAaBaORAaCa}
\left\{ \begin{array}{ll}
s^2 > \frac{(a-\mu)^2 (b-\mu)}{a+2b-3\mu} 
       & \text{if $\mu < a - \frac{|a-b|}{\sqrt{2}}$}
        \IEEEeqnarraynumspace\\
s^2 >  \frac{(a-\mu)^2 (b-a)}{a+b-2\mu}
       & \text{if $a-\frac{|a-b|}{\sqrt{2}} \leq \mu < \frac{a+b}{2}$}
        \IEEEeqnarraynumspace\\
\text{false,} & \text{if $\frac{a+b}{2} \leq \mu \leq b$} \IEEEeqnarraynumspace\\
%s^2 > \frac{(a-\mu)^2 (b-a)}{a+b-2\mu}, 
s^2 > \frac{(a-\mu)^2 (b-\mu)}{a+2b-3\mu} 
    & \text{if $\mu>b$.} 
    %\IEEEeqnarraynumspace
\end{array} \right.
\end{equation}

Likewise, we examine the terms in (\ref{eqn:ProofEMNoPos2}).
(Because of the assumption \mbox{$b>a$}, we cannot simply exchange $a$ and $b$
in (\ref{eqn:ProofEM:CondAaBaORAaCa}).)
The condition $A_b B_b > 0$ expands to
\begin{equation} \label{eqn:ProofEM:CondAbBb}
\left\{ \begin{array}{ll}
s^2 > \frac{(b-\mu)^2 (a-\mu)}{b+2a-3\mu} 
     & \text{if $\mu < a$ or $\mu>\frac{b+2a}{3}$} 
     \IEEEeqnarraynumspace\\
\text{false,} & \text{otherwise.}
            \IEEEeqnarraynumspace
\end{array} \right.
\end{equation}
The condition $A_b C_b>0$ expands to
\begin{equation} \label{eqn:ProofEM:CondAbCb}
\left\{ \begin{array}{ll}
\text{true,} & \text{if $\mu<a$}
            \IEEEeqnarraynumspace\\
%s^2 > \frac{(b-\mu)^2 (a-b)}{a+b-2\mu}, 
%     & \text{if $\mu < a$} 
%     \IEEEeqnarraynumspace\\
\text{false,} & \text{if $a \leq \mu \leq \frac{a+b}{2}$}
            \IEEEeqnarraynumspace\\
s^2 > \frac{(b-\mu)^2 (a-b)}{a+b-2\mu}
        & \text{if $\mu > \frac{a+b}{2}$.}
         \IEEEeqnarraynumspace
\end{array} \right.
\end{equation}
The functions
$\frac{(b-\mu)^2 (a-\mu)}{b+2a-3\mu}$
and $\frac{(b-\mu)^2 (a-b)}{a+b-2\mu}$
intersect at the points $\mu=b$ and
\begin{equation}
\mu = b \pm \frac{a-b}{\sqrt{2}}.
\end{equation}
We conclude that (\ref{eqn:ProofEMNoPos2}) is equivalent to
\begin{equation} \label{eqn:ProofEM:CondAbBbORAbCb}
\left\{ \begin{array}{ll}
s^2 > \frac{(b-\mu)^2 (a-\mu)}{b+2a-3\mu} 
    & \text{if $\mu<a$.} 
    \IEEEeqnarraynumspace\\
\text{false,} & \text{if $a \leq \mu \leq \frac{a+b}{2}$} \IEEEeqnarraynumspace\\
s^2 >  \frac{(b-\mu)^2 (a-b)}{a+b-2\mu}
       & \text{if $\frac{a+b}{2} < \mu \leq b + \frac{|a-b|}{\sqrt{2}}$}
        \IEEEeqnarraynumspace\\
s^2 > \frac{(b-\mu)^2 (a-\mu)}{b+2a-3\mu} 
       & \text{if $\mu > b + \frac{|a-b|}{\sqrt{2}}$.}
        \IEEEeqnarraynumspace
\end{array} \right.
\end{equation}

\begin{figure}
\centering
\includegraphics[width=\linewidth]{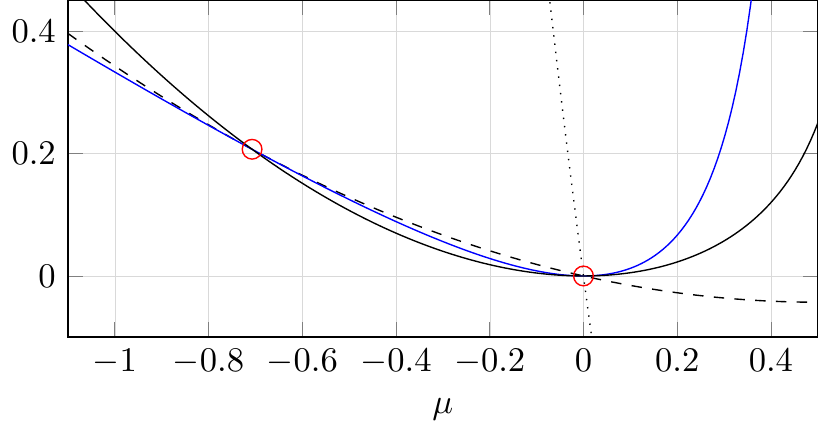}
\vspace{-3mm}

\caption{\label{fig:EMProof:curves}%
The functions in Table~\ref{table:EMProof:curves}
for $a=0$ and $b=1$.}
\end{figure}

We have thus expanded Lemma~\ref{lemma:ProofEM:NoPositiveRoots1} into
\begin{lemma}\label{lemma:ProofEM:NoPositiveRoots2}
Assume that (\ref{eqn:ProofEMDerSigmaaZeroSepDiscr}) holds.
Then (\ref{eqn:ProofEMDerSigmaaZeroSep}) has no positive real zero
if and only if (\ref{eqn:ProofEM:CondAaBaORAaCa}) holds,
and (\ref{eqn:ProofEMDerSigmabZeroSep}) has no positive real zero if and only if
(\ref{eqn:ProofEM:CondAbBbORAbCb}) holds.
\end{lemma}

For $\mu< \frac{a+b}{2}$, (\ref{eqn:ProofEM:CondAbBbORAbCb}) is stricter
than (\ref{eqn:ProofEM:CondAaBaORAaCa}), and for $\mu> \frac{a+b}{2}$,
(\ref{eqn:ProofEM:CondAaBaORAaCa}) is stricter than (\ref{eqn:ProofEM:CondAbBbORAbCb}).
This allows us to combine (\ref{eqn:ProofEM:CondAaBaORAaCa}) and (\ref{eqn:ProofEM:CondAbBbORAbCb})
into
\begin{lemma}\label{lemma:ProofEM:NoPositiveRoots3}
Assume that (\ref{eqn:ProofEMDerSigmaaZeroSepDiscr}) holds.
Then at least one of (\ref{eqn:ProofEMDerSigmaaZeroSep}) and (\ref{eqn:ProofEMDerSigmabZeroSep})
has no positive real zeros
if and only if
\begin{equation} \label{eqn:ProofEM:NoPositiveRoots3}
\left\{ \begin{array}{ll}
s^2 > \frac{(a-\mu)^2 (b-\mu)}{a+2b-3\mu} 
       & \text{if $\mu < a - \frac{|a-b|}{\sqrt{2}}$}
        \IEEEeqnarraynumspace\\
s^2 >  \frac{(a-\mu)^2 (b-a)}{a+b-2\mu}
       & \text{if $a-\frac{|a-b|}{\sqrt{2}} \leq \mu < \frac{a+b}{2}$}
        \IEEEeqnarraynumspace\\
s^2 >  \frac{(b-\mu)^2 (a-b)}{a+b-2\mu}
       & \text{if $\frac{a+b}{2} < \mu \leq b + \frac{|a-b|}{\sqrt{2}}$}
        \IEEEeqnarraynumspace\\
s^2 > \frac{(b-\mu)^2 (a-\mu)}{b+2a-3\mu} 
       & \text{if $\mu > b + \frac{|a-b|}{\sqrt{2}}$.}
        \IEEEeqnarraynumspace
\end{array} \right.
\end{equation}
\end{lemma}

Using Table~\ref{table:EMProof:curves},
Lemma~\ref{lemma:ProofEM:NoPositiveRoots3} and Lemma~\ref{lemma:ProofEM:ComplexZeros}
can be combined into
\begin{lemma}\label{lemma:ProofEM:NoPositiveRoots4}
At least one of (\ref{eqn:ProofEMDerSigmaaZeroSep}) and (\ref{eqn:ProofEMDerSigmabZeroSep})
has no positive real zeros if and only if 
\begin{equation} \label{eqn:ProofEM:NoPositiveRoots4}
\left\{ \begin{array}{ll}
s^2 > (3-\sqrt{8})(a-\mu)(b-\mu)
       & \text{if $\mu < a - \frac{|a-b|}{\sqrt{2}}$}
        \IEEEeqnarraynumspace\\
s^2 >  \frac{(a-\mu)^2 (b-a)}{a+b-2\mu}
       & \text{if $a-\frac{|a-b|}{\sqrt{2}} \leq \mu < \frac{a+b}{2}$}
        \IEEEeqnarraynumspace\\
s^2 >  \frac{(b-\mu)^2 (a-b)}{a+b-2\mu}
       & \text{if $\frac{a+b}{2} < \mu \leq b + \frac{|a-b|}{\sqrt{2}}$}
        \IEEEeqnarraynumspace\\
s^2 > (3-\sqrt{8})(a-\mu)(b-\mu)
       & \text{if $\mu > b + \frac{|a-b|}{\sqrt{2}}$.}
        \IEEEeqnarraynumspace
\end{array} \right.
\end{equation}
\end{lemma}

Finally, we note from Table~\ref{table:EMProof:curves}
that, for $\mu<(a+b)/2$, (\ref{eqn:s2Cond4a}) 
is implied by (\ref{eqn:ProofEM:NoPositiveRoots4}).
Likewise, for $\mu>(a+b)/2$,  
(\ref{eqn:s2Cond4b}) is also 
implied by (\ref{eqn:ProofEM:NoPositiveRoots4}),
which completes the proof.

\section*{Acknowledgment}

The authors would like to thank Gian Marti and Hampus Malmberg for 
valuable discussions.

\balance
\bibliographystyle{IEEEtran}
\bibliography{paper}

% that's all folks
\end{document}